\documentclass[12pt]{article}

\usepackage[top=1in,bottom=1in,right=1in,left=1in]{geometry}

\usepackage{amsmath, amssymb,bm}
\usepackage{array}
\usepackage{multirow}
\usepackage{multicol}
\usepackage{siunitx}
\usepackage{framed}
\usepackage{tabularx}
\usepackage{subcaption}
\usepackage[]{algorithm, algorithmic}
\usepackage{xcolor}
\usepackage{graphicx}
\usepackage[utf8]{inputenc}
\usepackage[noibid,authordate,natbib]{biblatex-chicago}
\usepackage{setspace}
\usepackage{tikz}
\usepackage{comment,wasysym}
\usepackage{listings}
\usepackage{mathtools}
\usepackage{enumitem}
\usepackage{xr}
\usepackage{hyperref}

\pgfdeclarelayer{bg}    
\pgfsetlayers{bg,main}  

\newcolumntype{L}[1]{>{\raggedright\let\newline\\\arraybackslash\hspace{0pt}}m{#1}}

\newcommand{\Gm}{\mathrm{Gamma}}

\newtheorem{theorem}{Theorem}
\newtheorem{lemma}{Lemma}
\newtheorem{definition}{Definition}
\newtheorem{coro}{Corollary}
\newcommand{\rank}{\mathrm{rank}}

\newcommand{\normone}[1]{\left\lVert#1\right\rVert}

\DeclareMathOperator*{\argmax}{arg\,max} 
\DeclarePairedDelimiter\floor{\lfloor}{\rfloor}
\title{Modelling Heterogeneity Using Bayesian Structured Sparsity}

\author{Max Goplerud\thanks{I thank participants at PolMeth 2019, EPSA 2019, New Faces in Political Methodology XI, MPSA 2019, Princeton's Quantitative Social Science Colloquium, Asian PolMeth 2019, APSA 2018, PolMeth 2018, and Text-As-Data 2018 for helpful comments. I thank the following people for comments on earlier versions of the paper: Naoki Egami, Shusei Eshima, June Hwang, Kosuke Imai, Gary King, Dean Knox, Shiro Kuriwaki, Naijia Liu, Ian Lundberg, Walter Mebane, Sun Young Park, Casey Petroff, Tyler Simko, Diana Stanescu, Brandon Stewart, Marc Ratkovic, Dustin Tingley, Jeremy Wachter, and Soichiro Yamauchi provided helpful feedback on earlier drafts of this paper. I thank the Alexander and Diviya Magaro Peer Pre-Review Program at Harvard's Institute for Quantitative Social Science for providing helpful feedback from an anonymous reviewer. Prior versions of this paper were known as ``Modelling Heterogeneity Using Complex Sparsity''. All remaining errors are my own.}}

\begin{document}
	\maketitle
\thispagestyle{empty}
	
\begin{abstract}

\noindent How to estimate heterogeneity, e.g. the effect of some variable differing across observations, is a key question in political science. Methods for doing so make simplifying assumptions about the underlying nature of the heterogeneity to draw reliable inferences. This paper allows a common way of simplifying complex phenomenon (placing observations with similar effects into discrete groups) to be integrated into regression analysis. The framework allows researchers to (i) use their prior knowledge to guide which groups are permissible and (ii) appropriately quantify uncertainty. The paper does this by extending work on ``structured sparsity'' from a traditional penalized likelihood approach to a Bayesian one by deriving new theoretical results and inferential techniques. It shows that this method outperforms state-of-the-art methods for estimating heterogeneous effects when the underlying heterogeneity is grouped and more effectively identifies groups of observations with different effects in observational data.\newline


\textbf{Key Words:} structured sparsity; generalized LASSO; Bayesian methods; heterogeneous effects; machine learning

\end{abstract}

\clearpage

\setcounter{page}{1}

\onehalfspacing

\section{Introduction}
Data analysis in social science faces a fundamental trade-off. On the one hand, there is a huge amount of richness and complexity in each observation. Yet, statistical models abstract away from this by making assumptions regarding the comparability of observations, the constant effects of some variables, functional form assumptions, and more. The choice of assumptions is not merely a technical decision but can have crucial substantive implications. Different choices may lead to different substantive conclusions, and all sets of assumptions imply particular relationships between the observations. Whenever possible, the choice of assumptions should be motivated by the researcher's substantive knowledge of the underlying question.

Consider a standard regression context. Perhaps the most fundamental question is to understand the effect that some independent variable has on the outcome. In doing so, the most common---and strongest---assumption would be that this variable has the same effect on all observations or that a single aggregate effect was substantively interesting. Straightforwardly including the variable linearly in the regression provides an estimate of this effect, as long as other critical assumptions are satisfied. Focusing on that quantity, however, may mask important sub-group heterogeneity that the researcher wishes to explore. A much more flexible approach to doing so would be to estimate a separate effect for each combination of the other covariates. Unfortunately, this approach typically leads to unacceptably noisy estimates of the sub-group effects. Research into simplifying the problem by assuming some stability in the underlying heterogeneity is often known as ``estimating heterogeneous treatment effects'' (e.g. \cite{imai2011heteffect,grimmer2017ensemble}).

Existing approaches to simplification often bet on the idea that the ``no heterogeneity'' assumption (i.e. a single effect) is approximately right: Estimates for sub-groups should be stabilized by being pulled towards some global or aggregate effect. This assumption appears in hierarchical models, sparse methods, and---more implicitly---many methods based on regression trees. A key goal of this paper is to suggest, however, that this assumption, while useful in many circumstances, does not always match how researchers intuitively and naturally make sense of a complex phenomenon.

Rather, this paper builds on a different intuition: Complex phenomena can be understood by classifying observations into a small number of \emph{groups} while being guided by our prior knowledge (if available). Such group creation is ubiquitous in political science research and appears whenever scholars create typologies or categorical variables. Explanations based on groups are often desirable because they also easily interpretable and explainable to others. Further, researchers are often in the position of having \emph{some} prior knowledge or belief about group structure (e.g. neighbors are connected), although they are not sufficiently confident to authoritatively classify observations. Thus, traditional methods of encoding and estimating groups (e.g. indicator variables or hierarchical models that allow for heterogeneity by group) are unsuitable as a core assumption of those approaches is that the groups are known \emph{ex ante}. On the other extreme, existing clustering algorithms are often unable to effectively integrate the researcher's prior beliefs on group membership and thus fail to effectively leverage valuable information from prior research.

This paper provides a framework to quantitatively encode this intuition: At its core is an existing method known as ``structured sparsity'' (\cite{huang2011structured,bach2012structured,chen2012proximal}). It starts from an agnostic scenario where there are many parameters representing the effect in each small unit (e.g. treatment-covariate profile combination). Before estimating the model, the researcher then decides which of these units \emph{might} be directly connected together in the same group---e.g. neighboring counties or profiles with common characteristics---based on their prior knowledge of the substantive question at hand. Finally, the model is estimated and the data determines whether two units should be fused together (i.e. given the \emph{same} effect). This occurs if (a) their estimated effects are close and (b) prior knowledge permits their combination. Groups thus emerge from this process where two units are given the same effect. The name ``structured sparsity'' comes from the fact that it modifies a traditional sparse approach (e.g. the LASSO; \cite{tibshirani1996lasso}) in the following crucial way: While sparsity seeks parsimony by encouraging many parameters to be \emph{zero} (``sparse''), structured sparsity encourages many parameters to be \emph{equal}---resulting in clustering.

Unfortunately, in its existing form, structured sparsity is not suitable for most social scientific research. There are two main limitations. First, it is difficult to quantify uncertainty in the estimated parameters as standard techniques of differentiating the objective or bootstrapping are not appropriate. Second, existing inferential techniques are limited to particular structures and/or a linear likelihood. This leaves an undesirable situation where researchers may think that particular choices of structure \emph{should} be used, but would be unable to fit a model that matches their theoretical belief. 

I address these concerns by creating a Bayesian formulation of structured sparsity. This allows simple inference and uncertainty quantification for many likelihoods and any set of prior beliefs (structure). In doing so, I generalize existing Bayesian research that is focused on particular structures (e.g. \cite{kyung2010penalized,betancourt2017network,tansey2017multiscale,faulkner2018locally}). The prior proposed to induce Bayesian structured sparsity is novel, and thus I provide new theoretical results on when the resulting posterior is proper as the prior is usually improper by design. Further, since the non-Bayesian structured sparse estimate is of special interest, the second major technical contribution of the paper provides a new algorithm for fast estimation using an Expectation Maximization algorithm (\cite{dempster1977EM}). This generalizes non-Bayesian estimation techniques that are also limited to specific structures or likelihoods (e.g. \cite{arnold2016fitting,zhu2017admm,chen2012proximal,tansey2017multiscale}).

The framework outlined in this paper is most applicable in the common scenario where the researcher has a large number of categorical or binary variables and wishes to simplify the complexity by creating groups. I explore two different settings to show its effectiveness: First, I use simulations to sharpen the intuition that existing methods perform poorly when the underlying heterogeneity is based around unknown \emph{groups}. Using a simple example where half of the units have an effect of one, half have an effect of negative one, but their membership is unknown, I show that many state-of-the-art methods (e.g. BART, LASSO, etc.) perform poorly---even losing to conventional methods such random effects. Structured sparsity outperforms both sets of methods, however, as it pools information more effectively by \emph{creating} groups of observations with the same effect. This creates both better performing and more easily interpretable results versus conventional methods. 

Second, I re-examine an experiment on the effect of politicians claiming credit for spending projects in their districts (\cite{grimmer2014credit}). Existing work has re-analyzed this study to look for heterogeneous effects across the types of treatment and respondent (\cite{grimmer2017ensemble}), and it provides a fruitful test of how structured sparsity's intuition based on groups compares to existing approaches. 

Substantively, I show that this method resolves a puzzling result in the initial analysis where (i) conservatives reacted more \emph{positively} to Planned Parenthood centers than moderates and (ii) conservatives reacted more \emph{negatively} to the construction of gun ranges than moderates for certain combinations of treatment effects. Further, I demonstrate that the preference for a project sponsored by a co-partisan \citep{grimmer2014credit} is concentrated amongst conservative Republicans and, to a lesser degree, liberal Democrats. I further provide evidence of the importance of the role of groups by putting structured sparsity into an ensemble with many state-of-the-art methods. It show that it gets substantial non-zero weight and is often the highest weighted model. This suggests that relying on groups captures something important and distinct from existing approaches.

\section{Heterogeneous Effects of Credit-Claiming}
\label{section:outline_het}

I begin with a motivating example about legislator credit-claiming by Grimmer and co-authors (\cite{grimmer2014credit,grimmer2017ensemble}). It experimentally tests a long-standing theory in American politics where voters are believed to increase support for their elected representatives if their representative can ``claim credit'' for  projects in their district such as building a road or a bridge (\cite{mayhew1974congress}). I specifically focus on an experiment that explores which types of projects may get more ``reward'' from voters and whether there are certain types of projects that voters dislike. Table~\ref{tab:creditclaiming} outlines the factorial-style experiment fielded on Mechanical Turk where voters are either shown a control message about no project (10\% of the sample) or a hypothetical project (\cite[pp. 97-105]{grimmer2014credit}; \cite{grimmer2017ensemble}). It yields 216 unique treatments.

\begin{table}[!ht]
	\caption{Credit-Claiming Experiment}
	\label{tab:creditclaiming}
	\begin{subtable}[t]{.45\textwidth}
	\caption{Description of Treatment Factors}
	\begin{itemize}
		\item Type: Roads, Police, Parks, Fire Department, Gun Range, Planned Parenthood
		\item Money: \$50 thousand or \$20 million
		\item Stage : Will Request, Requested, or Secured
		\item Sponsor: Republican or Democrat
		\item Co-Sponsor: None, Senate Republican, or Senate Democrat
	\end{itemize}
	\end{subtable}%
	\begin{subtable}[t]{0.55\textwidth}
		\caption{Sample Treatment}
		\begin{itemize}
			\item Representative X, Democrat, and Senator Y, a Republican, requested \$20 million to purchase safety equipment for local police. Rep. X said ``This money would help our brave police officers stay safe as they protect our property from criminals.''
			\item Representative Z, Democrat, secured \$50 thousand for medical equipment at the local planned parenthood. Rep. Z said ``This money will help provide state of the art care for women in our community.''
		\end{itemize}
	\end{subtable}
	\caption*{\footnotesize \emph{Note}: A brief description of each of the treatment factors is found above. Each treatment consists of one level of each of the factors combined together into a message.  Two examples are shown on the right. The full text of each treatment can be found in \citet[p. 99]{grimmer2014credit} or in \citet{grimmer2017ensemble}. \normalsize}
\end{table}

After having seen the project, the voters are asked whether ``they `approve or disapprove' of the way the fictitious representative `is performing (his/her) job in Congress'" \citep[p. 98]{grimmer2014credit}; the outcome is binarized for their analysis. \citet{grimmer2017ensemble} consider whether the effects of a particular treatment vary by respondent characteristics; for example, do Democrats prefer certain types of projects more than Republicans? They also explore whether certain aspects of the treatment have interactive effects.

Such an analysis of heterogeneous effects can be conceptualized in the following way. First, the researcher creates an ``expanded'' design matrix that includes many possible interactions between treatments and respondents. \citet{grimmer2017ensemble} include all pairwise interactions between the treatment factors (e.g. type-money, type-stage, money-stage, etc.) as well as interacting those pairwise interactions with respondent characteristics (e.g. ideology-type-money). This generates a problem for inference as there are only 1,074 responses but nearly 300 covariates in the design mentioned above. Thus, some sort of stabilization is needed to draw reliable inferences.

Existing approaches to heterogeneous effects typically propose methods that make some variant of the following assumption: the majority of deviations from an ``aggregate'' effect should be stabilized by pulling them towards zero. That is, they presume a model where a single global effect (i.e. ``no heterogeneity'') can well describe most observations and that deviations from that global effect are likely small for most units. This characterizes hierarchical models with random slopes (\cite{gelman2006multi}), sparse models (\cite{imai2013findit}), and many methods based on regression trees (e.g. BART, causal forests; \cite{hill2011bart,wager2018forests}).\footnote{Consider a tree placed on a set of parameters corresponding to a one-hot coding of membership into units $u$. A tree is grown by first deciding which unit $u$ should be pulled off into its own group. The next step decides whether a second unit $u'$ should form it own group. And so on. As the complexity of the tree is limited to prevent over-fitting, only a small number of units will be given a heterogeneous effect versus a global baseline containing most units. See Appendix~\ref{app:hetsims} for more discussion.} 

However, this paper suggests that a different approach based on \emph{groups} may be more fruitful. As the variables are categorical or binary (as is common in social science), a natural way to think about simplifying the estimates is creating groups of treatments with similar effects. For example, the main effect of roads and parks might be grouped and/or the interactions between party and fire departments may be fused. As the number of groups and their membership is unknown (unlike in hierarchical models), a key benefit of structured sparsity is to \emph{estimate} the groups in a data-driven but theory-guided way.

\section{Modelling Heterogeneity Using Structured Sparsity}

I focus in the remainder of the paper on structured sparsity as it is a flexible method for creating groups that can explicitly incorporate theoretical information about group structure (\cite{huang2011structured,bach2012structured,chen2012proximal}). I outline the approach in the context of a simplified version of the credit-claiming experiment with two treatment factors: project type (Gun Range [``Gun''], Planned Parenthood [``PP''], or Road) and amount of funding (\$50k or \$20million).\footnote{Alternative approaches based on clustering methods (e.g. \cite{bonhomme2015gfe,shen2015mixture,shahn2017mixture,shiraito2016treatment}) are useful in certain contexts but lack the ability to easily incorporate prior knowledge on which groups are permissible. For example, creating only geographically contiguous groups in a spatial setting is difficult for the cited methods, but trivial for structured sparsity. A secondary limitation is that most approaches require fixing the number of groups for any single run of the model.}

The basic procedure creates a large number of interactions defining a set of units with the same effect. Groups of units are fused together and given the same effect and thereby clusters emerge. Thus, unlike ``top-down'' clustering approaches such as finite mixture models, groups emerge from the (pairwise) comparison of units and thus it is a ``bottom-up'' approach to creating clusters. 

The researcher's prior knowledge is critical here in deciding which comparisons are permissible. It is often the case that researchers wish to rule out certain units being put together and thus wish to constrain the group formation process. These beliefs about which units are possibly connected together gives the ``structure'' to structured sparsity. Structure is often helpfully illustrated visually; Figure~\ref{fig:show_structures} begins by creating a network of the units or effects under consideration (project type / amount combinations) and then linking together nodes that \emph{might} be connected together.

\begin{figure}[!ht]
	\caption{Different Types of Structured Sparsity}
	\label{fig:show_structures}
	\begin{center}
	\begin{subfigure}[b]{0.3\textwidth}
		\caption{Agnostic}
		\centering
		\begin{tikzpicture}
		\foreach \x/\y/\alph/\name in {0/2/a/\$50k\\Gun, 0/4/b/\$20m\\Gun, 2/2/c/\$50k\\PP, 2/4/d/\$20m\\PP, 4/2/e/\$50k\\Road, 4/4/f/\$20m\\Road}{
			\node[circle, fill=gray!30, minimum width=1cm,draw,align=center,font=\footnotesize] (\alph) at (.9 * \y cm, .9 * \x cm) {\name}; }
		\begin{pgfonlayer}{bg}  
		\foreach \alpha in {a,b,c,d,e,f}%
		{%
			\foreach \alphb in {a,b,c,d,e,f}%
			{%
				\draw (\alpha) -- (\alphb);%
			}}
			\draw [bend left=45] (a) to (e);
			\draw [bend right=45] (b) to (f);
			
			\end{pgfonlayer}
		\end{tikzpicture}
	\end{subfigure}
	\begin{subfigure}[b]{0.3\textwidth}
		\caption{Lattice}
		\centering
		\begin{tikzpicture}
		\foreach \x/\y/\alph/\name in {0/2/a/\$50k\\Gun, 0/4/b/\$20m\\Gun, 2/2/c/\$50k\\PP, 2/4/d/\$20m\\PP, 4/2/e/\$50k\\Road, 4/4/f/\$20m\\Road}{
			\node[circle, fill=gray!30, minimum width=1cm,draw, align=center, font=\footnotesize] (\alph) at (1 * \y cm, .9 * \x cm) {\footnotesize \name}; 
		}
		\begin{pgfonlayer}{bg}  
		\draw (a) -- (c);
		\draw (a) -- (b);
		\draw [bend left=45] (a) to (e);
		\draw (c) -- (e);
		\draw (c) -- (d);
		\draw (b) -- (d);
		\draw [bend right=45] (b) to (f);
		\draw (d) -- (f);
		\draw (e) -- (f);
		\end{pgfonlayer}
		\end{tikzpicture}
	\end{subfigure}
	\begin{subfigure}[b]{0.3\textwidth}
		\caption{Priority}
		\centering
		\begin{tikzpicture}
		\foreach \x/\y/\alph/\name in {0/2/a/\$50k\\Gun, 0/4/b/\$20m\\Gun, 2/2/c/\$50k\\PP, 2/4/d/\$20m\\PP, 4/2/e/\$50k\\Road, 4/4/f/\$20m\\Road}{
			\node[circle, fill=gray!30, minimum width=1cm,draw, align=center, font=\footnotesize] (\alph) at (1 * \y cm, .9 * \x cm) {\footnotesize \name}; 
		}
		\begin{pgfonlayer}{bg}  
		\draw (a) -- (b);
		\draw (c) -- (d);
		\draw (e) -- (f);
		\end{pgfonlayer}
		\end{tikzpicture}
	\end{subfigure}
	\end{center}
	\caption*{\footnotesize \emph{Note}: Each figure shows a possible structure. The lines connecting each variable represent the possibility of exactly fusing the associated variables together when using structured sparsity. The left panel shows a fully connected structure (all groups are possible); the middle shows a ``lattice'' structure where two nodes are connected if and only if they share either project type or funding amount. The right panel show a structure where nodes are connected if they share project type.}
\end{figure}

Figure~\ref{fig:show_structures} shows three possible structures for this application. First, there is a fully connected or  ``agnostic'' structure (\ref{fig:show_structures}a); it encodes the minimal possible prior knowledge in that any two effects could be clustered together and thus any groups are possible. While it is the most flexible, it does not include any prior knowledge; it therefore may be less efficient than methods that incorporate structure. Further, it may form groups that are hard to interpret substantively given that there is no prior knowledge providing coherence to the groups. 

Some substantive knowledge can be added in multiple ways; a more permissive but still flexible structure resembles a ``lattice''. This would suggest that two units can be fused together if they share \emph{either} type of project \emph{or} amount. Figure~\ref{fig:show_structures}b shows the visual representation. Note that two units might still be in the same group if they are connected indirectly; for example, Roads-\$50k and Roads-\$20million are fused and Roads-\$20million and Gun Range-\$20million are fused together. Thus, Roads-\$50k and Gun Range-\$20 million are in the same group although they do not share any similar characteristics. It does, however, ensure that no group can contain both of them without an appropriate connecting pathway. This gives some coherence to the groups that emerge and may make them more interpretable. 

Finally, a quite restrictive structure is shown in the right panel (Figure~\ref{fig:show_structures}c). It allows only for groups \emph{within} one dimension of the heterogeneity, e.g. that groups can only be formed by fusing together units with the same type; it thus ``prioritizes'' a particular dimension of heterogeneity. These structures are appropriate when there is a strong prior belief that one dimension of heterogeneity dominates and the other is secondary. 

They also differ in their limiting case; if all connections bind for agnostic or lattice structures, then the model becomes one with a single effect---a model with no heterogeneous effects and a single treatment indicator. For the priority structure, by contrast, the limiting case is a model with one effect for each project type and thus equivalent to a model where only the factor for project type was included.

Formally, existing methods for structured sparsity operate using penalized maximum likelihood.  They begin by creating a vector of coefficients ($\bm{\beta}$) that represent the effect for each interactive combination; if there are $p$ effects, then $\bm{\beta}$ has $p$ levels where $\beta_i$ represents an indicator variable for observations that have treatment combination $i$ (e.g. Roads-\$50k). The un-penalized model can be estimated in the usual way, given a likelihood function $\ell$ that depends on the observed data $\bm{X}$ and outcomes $\bm{y}$.

\begin{equation}
\hat{\bm{\beta}}_{\mathrm{MLE}} = \argmax_{\bm{\beta}} \ell(\bm{\beta}; \bm{X}, \bm{y})
\label{eq:MLE}
\end{equation}

Structured sparsity is induced by adding a penalty on $\bm{\beta}$ to Equation~\ref{eq:MLE} that encourages elements of $\bm{\beta}$ to be set equal to each other (e.g. \cite{bondell2009anova,tibshirani2011general,gertheiss2010sparse,ma2017concave,tansey2017multiscale}). To make this happen, the penalty must have two properties; following work on (regular) sparsity inducing penalties \citep{fan2001lqa}, it should be zero (provide no penalty) if and only if the two elements are equal and it should be non-differentiable around the point where the two elements are equal. The simplest penalty to do this is based on the LASSO \citep{tibshirani1996lasso}. The LASSO is often used to set coefficients equal to zero, i.e. a penalized maximum likelihood estimate where many elements of $\bm{\beta}$ were set exactly to zero.

By contrast, \emph{structured} sparsity notes that the goal is not to set coefficients equal to zero, but rather equal to \emph{each other} to create clusters of distinct values. This intuition is formalized by placing a penalty on the difference or gap between two coefficients: $|\beta_i - \beta_j|$.~
Thus, one can think of the problem as deciding which differences to set (exactly) to zero; if two coefficients have no difference between them, they are fused together into a group. More generally, this form of structured sparsity can be induced by additively placing penalties on linear combinations of some element of $\bm{\beta}$. If $\bm{d}_k \in \mathbb{R}^{p}$, then a penalty that includes $|\bm{d}_k^T\bm{\beta}|$ encourages $\bm{d}_k^T\bm{\beta} = 0$ to hold at the optimum. 

In the case analyzed above, $\bm{d}_k$ would have one element equal to one (i.e. corresponding to $\beta_i$) and one corresponding to negative one (i.e. corresponding to $\beta_j$) and thus $\bm{d}_k^T\bm{\beta} = \beta_i - \beta_j$. For example, $|\beta_{\mathrm{Road/50\mathrm{k}}} - \beta_{\mathrm{Road/20\mathrm{mil}}}|$ would be used in all three structures as it encourages those two effects to be fused but $|\beta_{\mathrm{Road/50\mathrm{k}}} - \beta_{\mathrm{Gun/20\mathrm{mil}}}|$ would only appear in the agnostic structure as the two effects share neither attribute in common. By adding one restriction for each edge in Figure~\ref{fig:show_structures}, one can formalize each structure with a penalty that is the sum of the absolute value of the relevant restrictions, e.g. $|\bm{d}_1^T\bm{\beta}| + |\bm{d}_2^T\bm{\beta}| + \cdots$. The specific choice of $\bm{d}_k$ to create groups of coefficients is thus only one application of structured sparsity. 

A natural question is whether it is possible to cluster \emph{multiple} effects together---rather than merely pairs. In the case of the credit-claiming example, this could fuse together levels from one factor (e.g. Roads and Guns) if and only if their interactions with other factors are all close in value. This can be done by adding a penalty of the form $\sqrt{\bm{\beta}^T \bm{F}_\ell \bm{\beta}}$ with a careful choice of symmetric positive semi-definite $\bm{F}_\ell$ that generalizes existing work on the (overlapping) group LASSO (\cite{yuan2006group,kyung2010penalized,jacob2009overlap}).

Combining these two types of penalty leads to the general definition of structured sparsity in Equation~\ref{eq:genlasso_main}. It contains $K$ linear penalties and $L$ quadratic penalties that represent the researcher's prior belief about the structure of the underlying groups.

\begin{equation}
\hat{\bm{\beta}}_{\mathrm{SSparse}} = \arg\max_{\bm{\beta}} \ell(\bm{\beta}; \bm{X}, \bm{y}) - \lambda \left[\sum_{k=1}^K |\bm{d}_k^T\bm{\beta}| + \sum_{\ell=1}^{L} \sqrt{\bm{\beta}^T\bm{F}_\ell\bm{\beta}}\right]
\label{eq:genlasso_main}
\end{equation}

$\hat{\bm{\beta}}_{\mathrm{SSparse}}$ is, unfortunately, rather difficult to estimate as the penalty is not differentiable, and standard optimization methods will fail (\cite{tseng2001convergence}). Many different solutions have been proposed to solve this (e.g. \cite{chen2012proximal,arnold2016fitting,zhu2017admm,gaines2018admm,chen2012proximal,tansey2017multiscale}). The cited methods are limited, however, in that it does not appear that an existing method can handle (i) non-linear models and (ii) an arbitrary choice of structure (i.e. arbitrary $\bm{d}_k$ and $\bm{F}_\ell$).

Further difficulties arise in quantifying the uncertainty around $\hat{\bm{\beta}}_{\mathrm{SSparse}}$. This occurs because the non-differentiable penalty that induces sparsity complicates traditional methods for quantifying uncertainty. Both directly examining the Hessian of the (penalized) log-likelihood and the bootstrap are inappropriate in this setting. The non-differentiability of the penalty precludes the former and the sparsity of the estimates limits the latter as it is known that the bootstrap will fail to correctly quantify uncertainty in the standard sparse case (e.g. \cite{leeb2005oracle,kyung2010penalized}) and thus similar problems likely arise here.\footnote{Adapting other methods from machine learning (e.g. \cite{chernozhukov2018dml}) to quantify uncertainty is left for future research.}

\section{Bayesian Structured Sparsity}
\label{section:bayesian_ssp}

This paper proposes a different way to perform inference on $\bm{\beta}$ to resolve these problems: Bayesian inference. Existing research has considered particular choices of structure (e.g. \cite{park2008bayesian,kyung2010penalized,betancourt2017network,tansey2017multiscale,faulkner2018locally,pauger2018fusion}), but the general case with multiple linear and quadratic restrictions remains unexplored. Thus, it is important to analyze the theoretical properties of Bayesian structured sparsity as it appears to be a new prior in its general form.

This theoretical section has two parts; first, I explicitly derive conditions when the structured sparse prior and resulting posterior is proper. Second, I outline how the posterior can be sampled and how this leads to a novel (non-Bayesian) method for finding the solution to Equation~\ref{eq:genlasso_main}. The Supporting Information provides a number of additional theoretical results. Section~\ref{app:sub_global_local} outlines an extension to other methods of inducing sparsity (``global-local'' priors; \cite{polson2011shrink}); Section~\ref{app:calibrate_lambda} provides a discussion of how to choose the optimal regularization strength ($\lambda$) in a computationally efficient manner. 

\subsection{Theoretical Analysis of Bayesian Structured Sparsity}

Existing Bayesian work using particular structures does not examine conditions for posterior propriety. Most applications note the prior is improper and adjust it in some \emph{ad hoc} fashion to ensure propriety (e.g. \cite{betancourt2017network,faulkner2018locally}).
~This is an important issue to rectify as an improper posterior could result from an improper prior. As Bayesian algorithms will seemingly function properly in this case where the model is not well-defined (\cite{hobert1996gibbs}), this poses a danger to reliable inference. This section resolves the problem by deriving some testable conditions for posterior propriety. To begin, Equation~\ref{eq:genlasso_main} is formalized as a prior in Definition~\ref{def:ssp}.

\begin{definition}
	\label{def:ssp}
	A ``structured sparse'' prior on $\bm{\beta} \in \mathbb{R}^p$ with some regularization strength $\lambda > 0$ penalizes $K$ linear constraints ($\bm{d}_k$) and $L$ quadratic constraints ($\bm{F}_\ell$) on the parameters where $\bm{F}_\ell$ is symmetric and positive semi-definite. The kernel of the prior can be expressed as follows:
	
	$$p(\bm{\beta}) \propto \exp\left(-\lambda \left[\sum_{k=1}^K |\bm{d}_k^T\bm{\beta}| + \sum_{\ell=1}^L \sqrt{ \bm{\beta}^T\bm{F}_{\ell} \bm{\beta}}\right] \right)$$
	
	Further define $\bm{D}^T = [d_1, \cdots, d_K]^T$ and $\bar{\bm{D}}^T = [\bm{D}^T, \bm{F}_1, \cdots, \bm{F}_L]$. The prior is proper if the integral of the kernel is finite.
\end{definition}

All proofs are found in Appendix~\ref{app:proof_ssp} and involves rotating $\bm{\beta}$ such that a flat prior is induced on $p-\rank(\bar{\bm{D}})$ components of $\bm{\beta}$. This immediately leads to Theorem~\ref{thm:proper_prior}.

\begin{theorem}
	\label{thm:proper_prior}
	For $\lambda >0 $, a prior specified by Definition 1 is proper if and only if $\bar{\bm{D}}$ is full column rank.
\end{theorem}

Unfortunately, Theorem~\ref{thm:proper_prior} rarely holds in practice as many theoretically motivated choices of $\bar{\bm{D}}$ are not full rank. Thus, Theorem~\ref{thm:proper_posterior} derives two simple and easily testable conditions for posterior propriety. Proof of sufficiency leverages existing results by \citet{michalak2016proper} on partially flat priors and the orthogonal rotation of $\bm{\beta}$.

\begin{theorem}
	\label{thm:proper_posterior}
	Assume a model of the following form:
	\begin{itemize}
		\item Likelihood: $L(\bm{\eta} | \bm{y}) \equiv f(\bm{y} | \bm{\eta})$ where $\bm{\eta} = \bm{X}\bm{\beta}$. $\bm{X} \in \mathbb{R}^{N \times p}$ and $\bm{\beta} \in \mathbb{R}^{p}$. Further, assume that the likelihood is log-concave with respect to $\bm{\eta}$.
		\item Prior: $p(\bm{\beta}) \propto \exp\left(-\lambda \left[||\bm{D}\bm{\beta}||_1 + \sum_{\ell=1}^L \sqrt{\bm{\beta}^T\bm{F}_\ell\bm{\beta}}\right]\right)$ as in Definition~\ref{def:ssp}.
	\end{itemize}
	
	Consider the following conditions;
	\begin{enumerate}[label=(\alph*)]
		\item Augmented Design is Full Rank: $\rank\left(\left[\bm{X}^T~\bar{\bm{D}}^T\right]\right) = p$
		\item Unique MLE of Maximally Sparse Model: $\hat{\bm{\beta}}_{\mathcal{N}(\bar{\bm{D}})}$ exists and is unique.
		\begin{subequations}
			\begin{alignat*}{2} 
			\hat{\bm{\beta}}_{\mathcal{N}(\bar{\bm{D}})} &=  \argmax_{\bm{\beta} \in \mathbb{R}^p}~L(\bm{\eta} | \bm{y}) \quad \mathrm{s.t.} \quad \bm{D}\bm{\beta} = \bm{0},~\bm{\beta}^T\bm{F}_\ell\bm{\beta} = 0 \quad \forall \ell
			\end{alignat*}
		\end{subequations}
	\end{enumerate}
	(a) is necessary for posterior propriety; (b) is sufficient for posterior propriety.
\end{theorem}

Corollary~\ref{coro:glm} sharpens the results for the most popular likelihoods. 
 
\begin{coro}
	\label{coro:glm}
	If the likelihood is linear or multinomial with a standard link (e.g. logistic or probit), then Conditions (a) and (b) in Theorem~\ref{thm:proper_posterior} are jointly necessary and sufficient for posterior propriety.
\end{coro}

Conditions (a) and (b) have intuitive interpretations. Condition (a) encodes the Bayesian version of the usual ``full rank'' condition for maximum likelihood estimation to have a unique solution. In the Bayesian version, a design matrix ($\bm{X}$) that is not full rank---as is common in high dimensional cases---can still be associated with a proper posterior if either the prior is proper (i.e. $\rank(\bar{\bm{D}}) = p$) or if, together, they create an augmented design matrix that is full rank. Condition (b) considers the following limiting case; assume all restrictions were binding so the model was the maximally sparse one permitted \emph{given} a particular structure. If the MLE of that model is unique and exists, then the posterior must be proper. 

In the case of the agnostic and lattice structures discussed above, Condition (b) involves on checking whether a model with a single treatment indicator (i.e. no heterogeneity) has a unique and finite MLE. In a normal experimental set-up, this is trivially satisfied. In the priority (Type) structure, this involves checking whether a MLE with a six-leveled treatment for project type results in a unique and finite MLE. Again, this is usually satisfied.

Overall, Theorem~\ref{thm:proper_posterior} and Corollary~\ref{coro:glm} can be easily checked before estimation and thus ensure that researchers can safely use the proposed method.

\subsection{Estimating Models with Bayesian Structured Sparsity}

Given a proper posterior, estimation in a Bayesian framework typically seeks to sample from the posterior implied by Theorem~\ref{thm:proper_posterior}. Theorem~\ref{thm:gibbs} facilitates this using data augmentation. 

\begin{theorem}
	\label{thm:gibbs}
	Given a structured sparse prior with $\bm{D}$ and $\{\bm{F}_\ell\}_{\ell=1}^L$ such that the prior is proper, the following joint density preserves a marginal structured sparse prior on $\bm{\beta}$.
	\begin{alignat}{2}
	p\left(\bm{\beta}, \{\tau^2_k\}_{k=1}^K, \{\xi^2_\ell\}_{\ell=1}^L | \lambda\right) \propto \begin{split} \exp\left(-\frac{1}{2}\bm{\beta}^T\left[\sum_{k=1}^K \frac{\bm{d}_k\bm{d}_k^T}{\tau^2_k} + \sum_{\ell=1}^L \frac{\bm{F}_\ell}{\xi^2_\ell}\right]\bm{\beta}\right) \\ 
	\prod_{k=1}^K \frac{\exp(-\lambda^2/2 \cdot \tau^2_k)}{\sqrt{\tau^2_k}} \prod_{\ell=1}^L \frac{\exp(-\lambda^2/2 \cdot \xi^2_\ell)}{\sqrt{\xi^2_\ell}} \end{split}
	\end{alignat}
\end{theorem}

Appendix~\ref{app:EM_multinomial} shows that posterior can be sampled using a Gibbs Sampler and a linear likelihood. It further shows how data augmentation can be used to make binomial and multinomial outcomes easily tractable.

Additionally, recall that a motivating rationale for Bayesian estimation was the difficulty of finding a penalized MLE---corresponding to the posterior mode. Even if Bayesian estimation is preferred for many analyses, having the ability to find a penalized MLE is useful for prediction tasks involving cross-validation and calibration of the fully Bayesian model as discussed in Appendix~\ref{app:calibrate_lambda}. Appendix~\ref{app:EM_multinomial} notes how Theorem~\ref{thm:gibbs} immediately leads to a stable and tractable estimation method for a penalized MLE using Expectation Maximization (\cite{dempster1977EM}) that works for \emph{all} structures and many non-linear likelihoods. This extends a smaller literature that uses an EM approach to estimate sparse models (e.g. \cite{figueiredo2003adaptive,polson2011svm,ratkovic2017sparse}).

\section{Simulations on Estimating Heterogeneous Effects}
\label{section:het_sims}

I show the use of structured sparsity by first illustrating a claim earlier in the paper that existing methods fare poorly when the underlying heterogeneous effect is represented by \emph{groups} of distinct values. Table~\ref{tab:ssp_simenv} outlines a simple simulation environment to test this. 

It differs from existing simulations insofar as most units have an effect of either `1' or `-1' although which unit is in which group is unknown. This means that a global or aggregate effect is not representative of many units. This is a plausible case to describe real patterns of heterogeneous effects and thus it is important to understand how existing methods fare in this scenario. I also vary the amount of ``grouped'' structure by considering a more traditional set-up where many units have a treatment effect of zero.

\begin{table}[!ht]
\caption{Simulation Environment}
\label{tab:ssp_simenv}
\begin{tabular}{p{\linewidth}}
	\hline\hline
\end{tabular}
\begin{itemize}
	\item Parameters: $G$ units; $r$ observations per group. This implies $N = G \times r$ observations.
	\item Treatment Vector: Assume that $\mathcal{S}$ groups have an effect of `1', $\mathcal{S}$ have an effect of `-1' and $G - 2 \mathcal{S}$ have an effect of zero. Without loss of generality, assume the groups are ordered such that the first $\mathcal{S}$ have an effect of `-1' and the last $\mathcal{S}$ have an effect of `1'.
	$$\bm{\tau} = [\underbrace{-1, \cdots, -1}_{g \in \{1, 2, \cdots, \mathcal{S}\}}, \underbrace{0,\cdots, 0, \cdots, 0,}_{g \in \{\mathcal{S} + 1, \cdots, G - \mathcal{S}\}} \underbrace{1, \cdots, \cdots, 1}_{g \in \{G - \mathcal{S} + 1, \cdots, G\}}]$$
	\item Outcome: 
	\begin{itemize}
		\item The generative model is: $y_i = x_i + \tau_{g[i]} d_i + \epsilon_i; \quad x_i \sim N(0,1); \quad \epsilon_i \sim N(0,1)$
		\item Half of the units in each group are randomly assigned to treatment ($d_i = 1$).
	\end{itemize}
	\item Models: All models are correctly specified, i.e. include a control for $x_i$ and the possibility of estimating heterogeneous effects for each group $g$. For example, the LASSO model includes $x_i$ and interactions between $d_i$ and indicator variables for each group. Appendix~\ref{app:hetsims} outlines the full specification for each model.
\end{itemize}
\end{table}

\begin{figure}[!ht]
	\caption{Visualizing Selected Simulations}
	\label{fig:vis_select_het}
	\includegraphics[width=\textwidth]{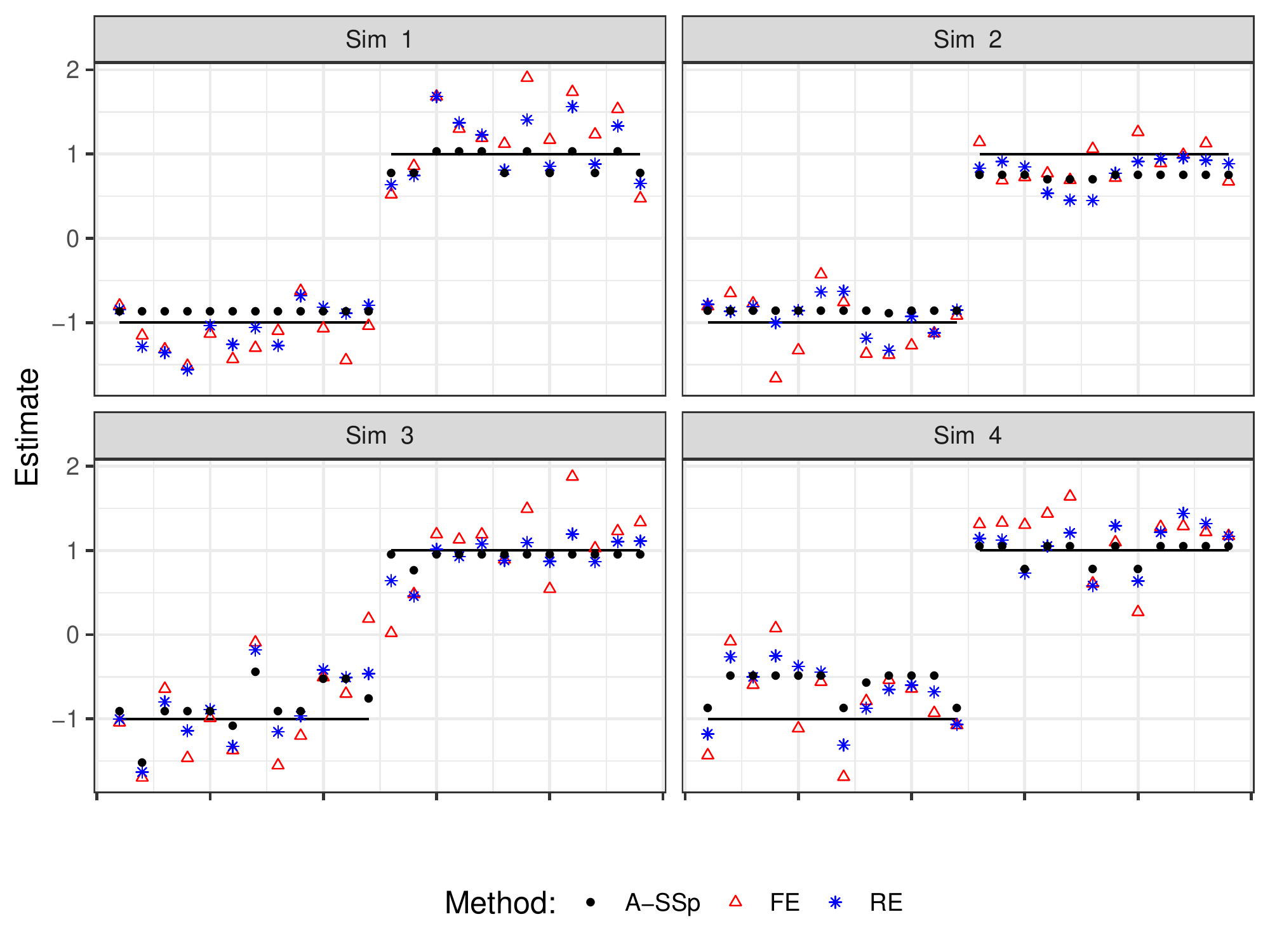}	
	\caption*{\footnotesize \emph{Note}: The estimated heterogeneous effect for each unit is shown; the units are arbitrarily to be in order of their true value. The three methods are fixed effects (FE), random effects (RE), and adaptive structured sparsity (A-SSp). The exact specification is described in Appendix~\ref{app:hetsims}. The true value for each point is shown with a thin black line. For clarity, the one unit with a true zero value is omitted.}
\end{figure}

To begin, I illustrate the results from a handful of simulations ($r = 20; G = 25$)to show the differences between structured sparsity and classical methods. Figure~\ref{fig:vis_select_het} presents the estimated heterogeneous effects from a fixed effects, random effects, and (adaptive) structured sparse model across four replications. Note that the points are \emph{arbitrarily} re-ordered such that the units with a negative effect are on the left and a positive effect are on the right. None of the estimated models know the true groups beforehand. In a model with perfect performance, all estimates would lie on the corresponding solid line.
	
The figure provides some intuition as to why stabilization is necessary and structured sparsity out-performs traditional methods: First, note that random effects performs better than fixed effects; fixed effects has around a 30\% worse RMSE averaged across the four plots in Figure~\ref{fig:vis_select_het}. However, neither approach meaningfully captures a ``group`` structure, although both usually show clear separation between the units with positive and negative effects. By contrast, structured sparsity clearly out-performs both noticeably and does a quite good job of recovering the unknown-but-underlying groups (fixed effects having a 95\% larger RMSE than structured sparsity; 44\% for random effects). While the grouping is not perfect, as expected given finite data, it allows for both a more interpretable result (i.e. the units are more often clearly separated) and more accuracy by more effectively pooling information. Its ability to recover the unknown-but-underlying groups shows its power and promise in this situation.

\color{black}

Table~\ref{tab:hetsims} examines this more systematically by showing the RMSE of the heterogeneous effects vs. the truth across eight key methods. It corroborates the benefit of using a regularization method based around groups. Consider first the upper panel where the underlying heterogeneity is indeed grouped (i.e. half of the heterogeneous effects are `1', half are `-1'). For all cases, structured sparsity performs the best. Even using a non-adaptive LASSO, this still outperforms all other methods although the adaptive LASSO weights give considerable performance gains. This confirms that it can effectively exploit group-structured heterogeneity when that well-describes the data generating process.
 
\begin{table}[!ht]
\caption{Estimating Heterogeneous Effects for $G = 25$ Units}
\label{tab:hetsims}
\begin{tabular}{c*{9}{c}}
\multicolumn{9}{c}{Mostly Grouped Effects ($\mathcal{S} = \floor{G/2}$)} \\
\hline\hline
$r$ & SSp & A-SSp & FE & RE & LASSO & ENet 1 & FindIt & BART \\
\hline
\multirow{2}{*}{10} & 0.411 & \textbf{0.401} & 0.630 & 0.427 & 0.565 & 0.562 & 0.443 & 0.970 \\
  & (0.008) & (\textbf{0.011}) & (0.009) & (0.007) & (0.007) & (0.006) & (0.006) & (0.001) \\
\multirow{2}{*}{20} & 0.282 & \textbf{0.229} & 0.445 & 0.313 & 0.407 & 0.416 & 0.315 & 0.920 \\
  & (0.005) & (\textbf{0.006}) & (0.006) & (0.005) & (0.005) & (0.005) & (0.005) & (0.002) \\
\multirow{2}{*}{50} & 0.180 & \textbf{0.115} & 0.277 & 0.202 & 0.262 & 0.275 & 0.196 & 0.535 \\
  & (0.003) & (\textbf{0.004}) & (0.004) & (0.003) & (0.004) & (0.004) & (0.003) & (0.007) \\
\multirow{2}{*}{100} & 0.134 & \textbf{0.078} & 0.197 & 0.145 & 0.189 & 0.199 & 0.143 & 0.294 \\
  & (0.002) & (\textbf{0.002}) & (0.002) & (0.002) & (0.002) & (0.002) & (0.002) & (0.003) \\
	
\hline\hline

\\

\multicolumn{9}{c}{Mostly Sparse Effects ($\mathcal{S} = \floor{G/4}$)} \\
\hline\hline
$r$ & SSp & A-SSp & FE & RE & LASSO & ENet 1 & FindIt & BART \\
\hline
\multirow{2}{*}{10} & \textbf{0.408} & 0.469 & 0.633 & 0.414 & 0.478 & 0.473 & 0.438 & 0.692 \\
  & (\textbf{0.006}) & (0.007) & (0.010) & (0.007) & (0.008) & (0.008) & (0.007) & (0.002) \\
\multirow{2}{*}{20} & \textbf{0.288} & 0.350 & 0.439 & 0.296 & 0.353 & 0.351 & 0.304 & 0.655 \\
  & (\textbf{0.004}) & (0.007) & (0.006) & (0.004) & (0.005) & (0.006) & (0.005) & (0.002) \\
\multirow{2}{*}{50} & 0.191 & \textbf{0.176} & 0.278 & 0.203 & 0.255 & 0.260 & 0.198 & 0.430 \\
  & (0.003) & (\textbf{0.005}) & (0.004) & (0.003) & (0.005) & (0.004) & (0.003) & (0.005) \\
\multirow{2}{*}{100} & 0.134 & \textbf{0.098} & 0.198 & 0.143 & 0.183 & 0.187 & 0.136 & 0.226 \\
  & (0.002) & (\textbf{0.003}) & (0.003) & (0.002) & (0.004) & (0.004) & (0.002) & (0.004) \\
	
\hline\hline
\end{tabular}
\caption*{\footnotesize \emph{Note}: The table reports the RMSE for the estimated heterogeneous effects versus the true values averaged across 100 simulations. The standard error is shown in parentheses. $r$ refers to the number of observations per group. The top panel shows the case where most effects are non-zero; the bottom shows mostly zero effects. See Table~\ref{tab:ssp_simenv} for details. Appendix~\ref{app:hetsims} details the methods and software used. The abbreviations stand for, in order, structured sparsity (SSp); an adaptive LASSO version (A-SSp); an interactive model (fixed effects - FE); a hierarchical model with random slopes (RE); LASSO; Elastic Net with $\alpha = 0.5$ (ENet1); FindIt \citep{imai2013findit}; Bayesian Additive Regression Trees (BART).}
\end{table}

Of the non-structured sparse methods, a simple hierarchical model with random slopes for treatment by group performs very well followed by traditional sparse methods and FindIt. The poor performance of tree-based methods (BART) suggests that they struggle in settings with categorical variables containing many levels (see Appendix~\ref{app:hetsims} for more discussion).

Consider the second case (the bottom panel; $\mathcal{S} = \floor{G/4}$) where the heterogeneity is much less grouped, i.e. half of the units have zero effect, one quarter have an effect of `1' and one quarter have an effect of `-1'. First, many traditional methods see improved performance, especially in the case of LASSO and Elastic Net for small $r$. This makes sense given that their underlying assumptions are more satisfied. Correspondingly, structured sparsity does worse. The adaptive version sees a noticeably degradation of performance, although it still remains the best performing method for moderately large $r$. Non-adaptive structured sparsity has stable performance and beats most other methods at all $r$, although the amount of improvement is certainly more limited. 

Taken together, the implications of these simulations is (i) structured sparsity can provide large gains in performance over traditional methods when the underlying heterogeneity is based on groups but (ii) has the ability to flexibly accommodate those scenarios leads to some cost in performance when the traditional methods have the correct assumption (mostly sparse effects). However, since structured sparsity can accommodate the traditional assumption as a special case, it still performs reasonably well. Given that it is not possible in real data to know which assumption is closer to the truth, structured sparsity is most effectively deployed when there is a reasonable suspicion that a grouped analysis might be effective to the question at hand but, unlike with random effects or models based on interactions, the groups cannot be precisely specified \emph{ex ante}.

Appendix~\ref{app:hetsims} provides additional results; first, it visualizes simulations in terms of percentage change in RMSE vs. adaptive structured sparsity to illustrate relative performance and adds four additional methods (SVM, random forest, ridge regression, another elastic net). Re-interpreting Table~\ref{tab:hetsims} in this light confirms the substantively size of the improvement: In the mostly grouped case, random effects and FindIt when $r> 10$ RMSEs that are often 40\% larger than adaptive structured sparsity and much worse (e.g. 80-100\%) when $r$ is large. In the mostly sparse case, the magnitude of improvements is smaller (e.g. only 20-50\% when $r$ is large) and that it is sometimes about 10\% worse than existing methods.

Second, I examine results with binary outcomes. Structured sparsity remains the best method in the mostly grouped case (although its magnitude of improvement is smaller). Its performance also degrades somewhat more sharply in the non-grouped case although for large $r$ consistently beats competitor methods by small but distinguishable margins.

Third, I vary the number of units to range from five to one-hundred. Except for the case of very few units ($G = 5$), structured sparsity continues to perform well. To summarize the eighty simulation environments (e.g. outcome type, $r$, $G$, and $\mathcal{S}$ combinations) in brief, structured sparsity and its adaptive analogue are in the top-three methods in 74 and 60 cases, respectively.

\section{Uncovering Heterogeneous Effects of Credit-Claiming}
\label{section:est_het}

As discussed previously, the experimental design in \citet{grimmer2014credit} contains five treatment factors and two variables (party identification and ideology) for respondent-specific heterogeneity. The survey contains only 1,074 observations and thus the ability to look for all possible combinations of treatment and respondent is significantly constrained. The existing literature suggests that the type of project is critically important (\cite{grimmer2014credit,grimmer2017ensemble}), and thus I generate all pairwise interactions between project type and each other type of treatment as well as the interactions with the respondent characteristics. The formula to generate the design is expressed below.

\begin{lstlisting}
~ Type * (Money + Stage + Sponsor + Co-Sponsor) * (Party + Ideology)
\end{lstlisting}

This leads to a design with 538 parameters and an intercept.\footnote{This is larger than $p = 280$ used in \citet{grimmer2017ensemble}'s as structured sparsity does not require the specification of an (arbitrary) baseline category for each treatment that may affect performance of, say, LASSO-based methods.} A key part of structured sparsity is using one's theory to decide how these should be possibly grouped together. I consider three possible strategies each reflecting increasingly strong theoretical knowledge. Table~\ref{tab:ensemble_structures} provides a brief summary as they are described in detail in Section~\ref{section:outline_het}.

\begin{table}[!ht]
	\caption{Description and Performance of Structures For Credit-Claiming Experiment}	
	\label{tab:ensemble_structures}
	\begin{center}
	\begin{tabular}{lllll}
		\hline\hline
		Name & Description & Limiting Case & WAIC & CV\\
		Agnostic & Connect all & Single treatment & 1294.7 & 0.451 \\
Lattice & Connect if they share any attribute & Single treatment & 1295.3 & 0.451 \\
Priority & Connect to project ``type'' & Single factor & 1309.8 & 0.457 \\

		\hline\hline
	\end{tabular}
	\end{center}
	\caption*{\footnotesize \emph{Note}: This table briefly describes each of three structures used in the analysis. The final columns (WAIC and CV) are measure of fit discussed in the main text; smaller values indicate better performance.}
\end{table}

The agnostic and lattice structures are generated straightforwardly. As noted above, their limiting case as $\lambda \to \infty$ is a model with only a single treatment indicator (i.e. no effect heterogeneity). The third structure (``priority'') uses the prior research on this topic to focus on the role of project type. It connects together all effects that contain the same type of project (e.g. Roads, Roads-\$50k, Roads-\$50k-Democrat). For terms without a project type, they are connected based on sharing the same treatment factor (e.g. \$50k, \$50k-Democrat, \$20million). Controls for respondent party and ideology are also included additively. Its limiting case is where only the factor for project type was included alongside the respondent controls. Appendix~\ref{app:grimmer} provides details on the calibration of $\lambda$ and posterior diagnostics.\footnote{I run the model for 4 chains with 5,000 samples after 5,000 burnin. All Gelman-Rubin diagnostics on $\bm{\beta}$ are good; there is sometimes poor mixing on $\lambda$. I show that fixing $\lambda$ at an optimal value returns nearly identical results.}

To choose between the structures, one can adopt either a theoretical or data-driven approach. My theoretical prior is to use the lattice structure as that provides some coherence to the groups while also not necessarily prioritizing project type to the exclusion of all others. A data-driven approach compares the fit of the models while noting that they may have different complexities.

Table~\ref{tab:ensemble_structures} reports two criteria. First, the Widely Applicable Information Criterion (WAIC; \cite{gelman2013understanding}) is a technique for comparing non-nested Bayesian models and is designed to approximate cross-validation. It can be interpreted like a standard information criterion (AIC or BIC) where smaller indicates a better fit. The evidence from this statistic suggests that the agnostic and lattice structures out-perform the priority structure but otherwise are similar. Second, using the fast EM algorithm and the (approximate) AIC, I conduct 20-fold cross-validation (used again in Section~\ref{section:ensemble}) and report the cross-validated root mean squared error. This statistic suggests nearly identical performance between the agnostic and lattice structures although their individual predictions are not perfectly correlated.

Thus, since theory suggests the lattice structure and the data-driven approach is roughly indifferent, I report results using the lattice structure. As there is always researcher discretion over the choice of structure, it is important to examine sensitivity to this. Appendix~\ref{app:grimmer} compares the estimated heterogeneous effects for each treatment-respondent combination. The agnostic and lattice estimates are nearly perfectly correlated ($\rho = 0.99$), while the priority structure recovers a somewhat distinct pattern ($\rho = 0.95$).

\subsection{Interpreting High-Dimensional Heterogeneity}

Interpretation of the estimated heterogeneity is complicated as there are nearly 2,000 treatment/respondent effects that can be predicted from the fitted model. For complex models such as random forests or ensembles, initial exploration often begins by calculating the types of profiles that have the largest effects and looking for commonalities.

Structured sparsity as well as other LASSO-based methods have the useful property of allowing a different initial exploratory step as the model itself returns parameter estimates interpretable (roughly) as in a generalized linear model. It is sensible to run the fast EM model to find the penalized MLE and tune $\lambda$ as noted earlier. The posterior mode will contain exact groups and it is possible to quickly ascertain which effects appear to be meaningfully distinct. Table~\ref{tab:largest_EM} does this by reporting the largest parameter estimates. It notes that, as is common for many applications, there is a large group that contains most of the parameters (estimated at 0.0486) with the others being located in mostly singleton clusters. The group-based stabilization occurs by regularizing effects to be close to their \emph{neighbors} versus being pulled towards zero as is traditional.

\begin{table}[!ht]
	\caption{Largest Parameter Estimates at Posterior Mode}
	\label{tab:largest_EM}
	\begin{subtable}[t]{0.5\textwidth}
		\caption{Most Positive}
		\begin{tabular}{p{6.5cm}l}
			\hline\hline
			Type(PP) * Ideology(Lib) & 0.72 \\
Sponsor(Rep) * Party(Rep) & 0.56 \\
Type(Gun) * Ideology(Cons) & 0.46 \\
Stage(Secure) * Ideology(Lib) & 0.26 \\
Money(\$50thousand) * Party(Rep) & 0.25 \\
Party(Dem) & 0.24 \\
Type(Roads) & 0.24 \\
Type(Roads) * Sponsor(Rep) & 0.24 \\
Type(Roads) * Co-Sponsor(Dem) & 0.24 \\
Type(Fire) & 0.23 \\

			\\ \hline\hline
		\end{tabular}
	\end{subtable}%
	\begin{subtable}[t]{0.5\textwidth}
		\caption{Most Negative}
		\begin{tabular}{L{6.5cm}l}
			\hline\hline
			Type(Gun) & -1.52 \\
Type(Gun) * Party(Dem) & -1.14 \\
Type(Parks) * Co-Sponsor(Dem) & -0.53 \\
Type(PP) * Ideology(Cons) & -0.46 \\
Type(Gun) * Ideology(Lib) & -0.37 \\
Co-Sponsor(Rep) * Ideology(Mod) & -0.33 \\
Type(Police) * Sponsor(Rep) * Ideology(Lib) & -0.24 \\
Stage(Request) & -0.14 \\
Sponsor(Dem) * Ideology(Cons) & -0.11 \\
Sponsor(Dem) * Party(Rep) & -0.09 \\

			\hline\hline
		\end{tabular}
	\end{subtable}
	\caption*{\footnotesize \emph{Note}: 487 of the 538 parameters are grouped together at 0.0486. The top ten largest and smallest parameter estimates are shown above where `` * '' indicates an interaction. Planned Parenthood is abbreviated to PP.}
\end{table}

It clearly illustrates that certain types of interactions appear to be driving the estimated heterogeneity. As noted in prior research (\cite{grimmer2017ensemble}), respondent characteristics (party and ideology) appear very important in moderating the effects of the credit-claiming proposals. Further, it immediately illustrates the substantive point where the \emph{combination} of a co-partisan sponsor and the respondent's party or ideology is highly relevant. I examine each of these in turn.

First, the structured sparse approach resolves an odd feature of the results presented in \citet[p. 428]{grimmer2017ensemble}: They examine treatment effect by project type, amount of funding or status of project, and respondent ideology. Close inspection reveals that certain types of treatments have the following effect pattern: Conservatives and liberals have more extreme treatment effects in the \emph{same} direction than moderates. While this could be true in theory, the presented results are implausible.

Specifically, it suggests that when reviewing proposals to claim credit for Planned Parenthood, both conservatives and liberals have a \emph{more positive} response than moderates. A similar problem occurs for proposals to create gun ranges where conservatives and liberals are seen to have a \emph{more negative} response than moderates. Given the polarization of American politics around these issues, neither estimated effect seems likely to generalize to the broader population as the prior expectation would be a (weak) ordering of effects by ideology.

This problem does not appear in their analysis of marginal effect of project type by party (\cite[p. 427]{grimmer2017ensemble}) but arises in both analyses where interactive treatment effects are considered and thus may be a ``paradox'' resulting from marginalization.
~This could also be statistical noise, but the ensemble approach used in that paper cannot speak to this point.

\begin{figure}[!ht]
	\caption{Average Treatment Effects by Project Type, Secondary Treatment, and Ideology}
	\label{fig:treatment_ideology}
	\includegraphics[width=0.9\textwidth]{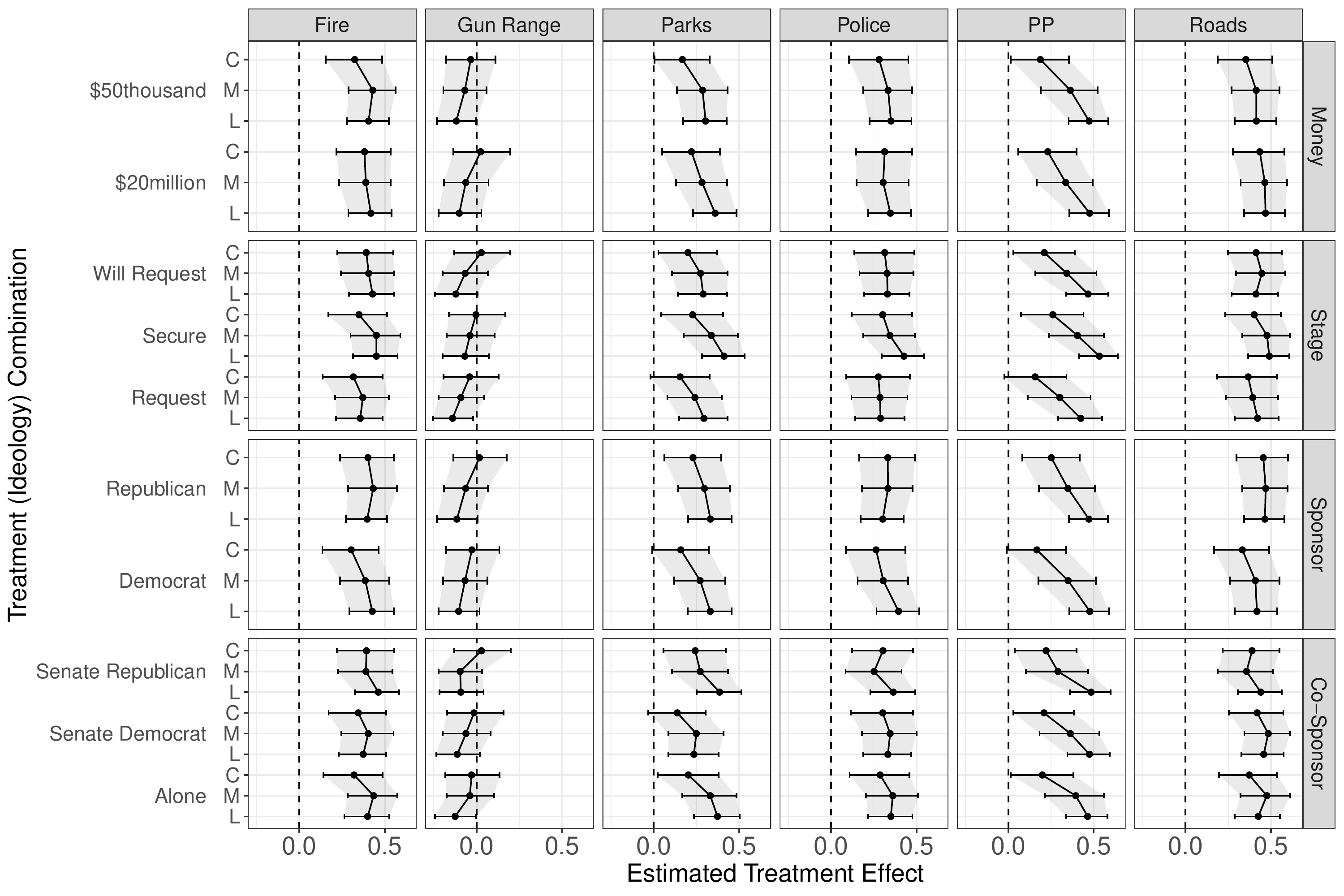}
	\caption*{\footnotesize \emph{Note}: This figure shows the treatment effect for project type (vertical panels) and other treatment factor (horizontal panel) by ideology of the respondent. `C', `M', and `L' denote Conservative, Moderate, and Liberal respondents, respectively. PP stands for Planned Parenthood. The 90\% credible interval is reported.}
\end{figure}

Figure~\ref{fig:treatment_ideology} explores this using structured sparsity. After calculating the analogous quantity, the point estimates for Planned Parenthood and Gun Ranges are sensible.\footnote{Formally, fix ideology, project type and one other treatment factor at say (liberal, Fire, and \$50 thousand). Find all heterogeneous effects with that combination of characteristics and average across them.} Across all other types of treatment factors, the posterior medians have a sensible ideological ordering of liberals, moderates, and conservatives for both Planned Parenthood and gun ranges. The interactive treatment effects thus recover the pattern in the marginal effects. Looking across other issues, there also appears to be little evidence of moderates having more extreme treatment effects that is distinguishable from statistical noise.

\begin{figure}[!ht]
	\caption{Average Marginal Effect of Republican Sponsor by Party and Ideology}
	\label{fig:AME_copart}
	\includegraphics[width=\textwidth]{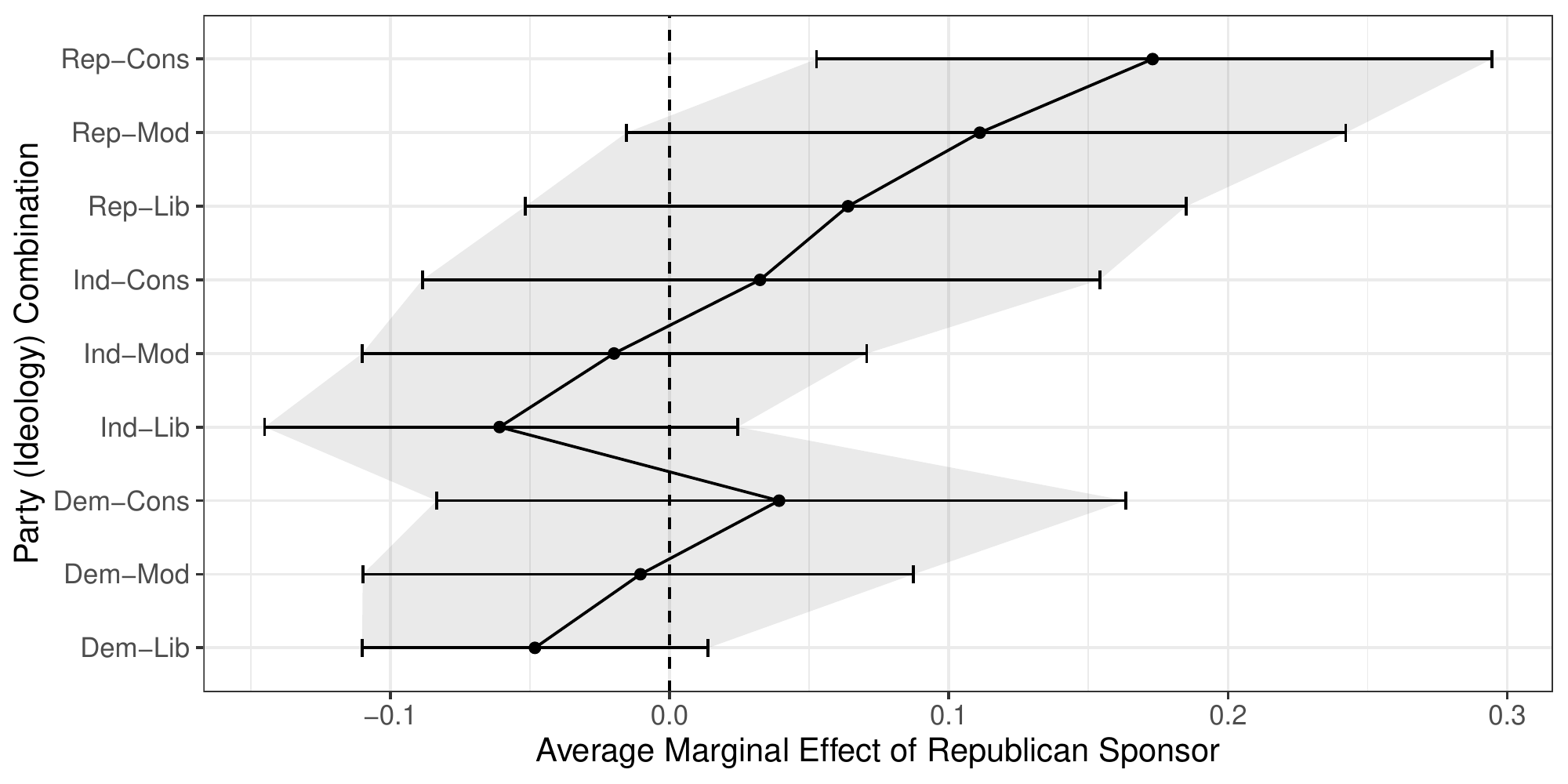}
	\caption*{\footnotesize \emph{Note}: This figure shows the difference between having a Republican and Democratic sponsor of the project by party and ideology combinations. The 90\% credible interval is reported.}
\end{figure}

The second substantive finding from structured sparsity concerns the possibility of an interactive effect between the party of the representative proposing the project and the respondent. In their original analysis, \citet[pp. 104-105]{grimmer2014credit} examine the role of this effect in a model that does not estimate complex heterogeneity. They find evidence for the role of a larger effect when the respondent has a co-partisan sponsor, although the difference does not always rise to the conventional level of statistical significance. I re-examine this and focus on the differential co-partisan effect by the party and ideology of the respondent as suggested by Table~\ref{tab:largest_EM}. Figure~\ref{fig:AME_copart} tests this by calculating the conditional average marginal effect of having a Republican sponsor vs. a Democratic sponsor.\footnote{Formally, fix party and ideology at some combination, say Moderate Independent. Next, calculate the difference in treatment effect between having a Republican and Democratic sponsor for every combination of other treatment factors and that demographic profile. Averaging across those combinations returns the average marginal effect.}

It confirms the initial results from analyzing the posterior mode; conservative Republicans show a noticeable preference for proposals from co-partisan representatives (Republicans) over Democratic proposers that is statistically distinguishable from zero. Liberal Democrats show the opposite bias (preferring Democratic sponsors), although the credible interval contains zero. This provides an interesting qualification of the results from \citet{grimmer2014credit}: There is a statistically distinguishably larger effect for co-partisans evaluating a representative's attempt to claim credit, although it is concentrated in respondents who share the dominant ideological beliefs of their party. 

\section{Structured Sparsity in an Ensemble}
\label{section:ensemble}

A final test for structured sparsity is how well it performs against other methods for detecting heterogeneous effects on real data. \citet{grimmer2017ensemble} combine a variety of methods using the idea of ``super learning'' (\cite{van2007super}). This approach fits a variety of methods using $K$-fold cross-validation and then examines how well each method's out-of-sample predictions correspond to the truth. It gives weights to each method that are non-negative and sum to one. Higher weights indicate that the model is used more heavily in the final predictive model and thus is seen to be better at some combination of predicting the outcome and/or being distinct from other methods.

Table~\ref{tab:ensemble} shows how structured sparsity fares using a 20-fold split of the data. I replicate the exact collection of models in the ensemble in \citet{grimmer2017ensemble} and add structured sparse models from Table~\ref{tab:ensemble_structures}. I also provide a ``harder'' test by including additional methods (Columns 3-4).\footnote{Note that a random effects model cannot be easily fit here. It would require around \emph{thirty} random effects to capture the proposed interactions (seven for the main effects plus around twenty for the interactions) and thus is prohibitively expensive. Thus, while Section~\ref{section:het_sims} suggests that random effects can perform well, an additional benefit of structured sparsity is its scalability.}

\begin{table}[!ht]
	\caption{Ensemble Analysis}
	\label{tab:ensemble}
	\begin{center}
	\begin{tabular}{l*{4}c}
		\hline\hline
		& \multicolumn{2}{c|}{Initial Specification} & \multicolumn{2}{c}{Structured Sparsity} \\
		Model & (1) & (2) & (3) & (4) \\
		\hline
		LASSO & 0.727 & 0.534 & 0.246 & 0.251  \\
ENet1 & - & - & - & -  \\
ENet2 & - & - & - & -  \\
BayesGLM & - & - & - & -  \\
FindIt & 0.273 & - & - & -  \\
KRLS & - & - & - & -  \\
SVM & - & - & - & -  \\
Boost &  & - &  & -  \\
BART &  & - &  & -  \\
RF &  & 0.107 &  & 0.013  \\
GLM &  & 0.359 &  & -  \\
SSp (A) &  &  & 0.281 & 0.285  \\
SSp (L) &  &  & 0.473 & 0.452  \\
SSp (P) &  &  & - & -  \\
	
		\hline\hline	
	\end{tabular}
	\caption*{\footnotesize \emph{Note}: Appendix~\ref{app:hetsims} describes the formula and software to fit all models. In addition to models in Table~\ref{tab:hetsims}, this table includes an Elastic Net with $\alpha = 0.25$ (ENet2), boosting (Boost), Kernel Regularized Least Squares (KRLS); a model with a weakly informative prior (BayesGLM), a support vector machine and a simple logistic regression with each treatment factor and covariate added linearly (GLM). SSp(A), SSp(L), SSp(P) refer to the agnostic, lattice, and priority structures described in Table~\ref{tab:ensemble_structures} respectively. The ensemble is constructed using 20-fold cross-validation. $-$ indicates a model was given zero weight.}
	\end{center}
\end{table}

It returns strong support for structured sparsity; in both settings when it is added, two structured sparse models (agnostic and lattice) receive substantial non-zero weight---comprising nearly three-quarters of the total weight in the ensemble. The fact that both models get some weight suggests they are each contributing something distinct and useful to the ensemble. This again suggests the utility of examining multiple structures. It further suggests that while structured sparsity is not the optimal model in all cases, it is worth including in ensemble analyses as it can cover the possibility of there being grouped heterogeneity in the underlying data.


\section{Conclusion}

This paper began by noting that, when modelling heterogeneity, it is often necessary to make simplifying assumptions to obtain relatively precise estimates. A vast array of methods provide different sets of assumptions to do this, mostly focused around stabilizing estimates around some global or aggregate effect. The fundamental contribution of this paper is allowing researchers to use a different assumption: Stabilize effects by creating data-driven groups guided by prior knowledge. Given that the creation of groups is fundamental to how researchers create concepts and make sense of the world, this method is applicable to many research questions across subfields---especially when the proposed model includes many binary or categorical variables.

To do this, existing methods for structured sparsity needed substantial modification to be suitable for social scientific research. The paper developed a Bayesian formulation of structured sparsity that allowed for the quantification of uncertainty in the estimated effects as well as tractable inference for non-linear models with arbitrary penalties. After deriving novel theoretical results about the propriety of the resulting posterior and new inferential techniques, I showed that structured sparsity performed well in two contexts. 

First, I showed that using a simple simulation where the underlying heterogeneous effects fell mostly into two distinct groups, state-of-the-art methods struggled to accurately estimate the heterogeneous effects. Their approach of shrinking aggressively towards a global effect failed to do well insofar as that global effect did not represent many groups. By contrast, structured sparsity---as well as traditional methods such as hierarchical models---performed well. The key benefit of structured sparsity over traditional methods such as random effects is that it can \emph{estimate} the group memberships, while being guided by theory, rather than requiring them to be exactly specified \emph{ex ante}.

Second, I applied Bayesian structured sparsity to a recent experiment on credit-claiming and drew out two new substantive implications. First, an unusual published result of moderates being more ideologically extreme for certain combinations of treatments  did not occur when using structured sparsity. For salient policies such as building gun ranges or Planned Parenthood clinics, structure sparsity estimated marginal as well as interactive treatment effects that agreed with the dominant understanding of American politics. 

Guided by the interpretable output from structured sparsity, I examined the claim that respondents react more favorably to policies proposed by a co-partisan representative (\cite{grimmer2014credit}). I also found qualified evidence for this but noted that the effect appears concentrated in conservative Republicans and liberal Democrats, although only the former shows effects that were statistically distinguishable from zero. Finally, I compared structured sparsity against a wide array of state-of-the-art methods in an ensemble, I found that structured sparsity had the best cross-validated error and received significant non-zero weight in an ensemble of methods stacked together.

Overall, the paper demonstrates a dual role to structured sparsity; first, if a single model is desired and one believes that \emph{groups} may represent a reasonable way of representing the underlying heterogeneity, it is a flexible general purpose tool for uncovering easily interpretable heterogeneity while quantifying uncertainty. Second, if one wishes to rely on a collection of models, its inclusion into an ensemble can account for the possibility of grouped heterogeneity and thus improve the performance of this power technique.

\printbibliography

\appendix

\setcounter{theorem}{0}
\renewcommand{\thetheorem}{\thesection.\arabic{theorem}}

\clearpage

\section{Proof of Theorems on Structured Sparsity}
\label{app:proof_ssp}

I note a number of standard results from linear algebra that are referenced in the proofs.

\begin{enumerate}[label=R\arabic*]
	\item For any $\bm{x} \in \mathbb{R}^{p}$, $||\bm{x}||_2 \leq ||\bm{x}||_1 \leq \sqrt{p} ||\bm{x}||_2 \leq \sqrt{p} ||\bm{x}||_1$
	\item Any matrix $\bm{D} \in \mathbb{R}^{n \times p}$ admits a singular value decomposition (SVD) of the following form where $\bm{\Sigma}$ is a diagonal matrix with positive entries of size $\rank(\bm{D}) = m$ and $\bm{U}$ and $\bm{V}$ are orthogonal matrices of size $n \times n$ and $p \times p$, respectively.
	\begin{equation}
	\bm{D} = \bm{U} \left(\begin{array}{ll} \bm{\Sigma} & \bm{0}_{m \times p - m} \\ \bm{0}_{n - m \times m} & \bm{0}_{n - m \times p - m} \end{array}\right) \bm{V}^T
	\end{equation}
	A ``thin'' SVD is defined as $\bm{D} = \bm{U}_1 \bm{\Sigma} \bm{V}_1^T$ where the subscript `1' denotes taking the first $m$ columns of a matrix.
	\item Assume that $\bm{X} \in \mathbb{R}^{n \times p}$ and $\bm{D} \in \mathbb{R}^{m \times p}$. Denote $\mathcal{N}(\bm{A})$ as the nullspace of $\bm{A}$ and $\mathcal{B}_{\bm{A}}$ as a basis for this nullspace. Define the set $\mathcal{S} \subseteq \mathbb{R}^p$ such that $\mathcal{S} = \{s: \bm{X}\bm{s} = \bm{0} ~\mathrm{and}~  \bm{D} \bm{s} = \bm{0}\}$. The following conditions are all equivalent:
	\begin{enumerate}
		\item $\mathcal{S} = \{\bm{0}\}$, i.e. the only member of $\mathcal{S}$ is the zero vector $\bm{0}$.
		\item $\mathcal{N}(\bm{X}) \cap \mathcal{N}(\bm{D}) = \{\bm{0}\} $
		\item $\rank\left(\left[\bm{X}^T~\bm{D}^T\right]\right) = p$
		\item $\rank(\bm{X} \mathcal{B}_{\bm{D}} ) = p - \rank(\bm{D})$
	\end{enumerate}
	\item Consider the following optimization problem where $f: \mathbb{R}^p \to \mathbb{R}$ and $\bm{A} \in \mathbb{R}^{K \times p}$.
	
	\begin{equation}
	\label{app_eq:constrained}
	\bm{x}^* = \argmax_{\bm{x} \in \mathbb{R}^p} f(\bm{x}) \quad \mathrm{s.t.} \quad \bm{A}\bm{x} = \bm{0}
	\end{equation}
	
	If $\mathcal{B}_{\bm{A}}$ is a basis for the nullspace of $\bm{A}$, i.e. consisting of $p - \rank(\bm{A})$ linearly independent vectors of length $p$, derive the following unconstrained optimization problem:
	
	\begin{equation}
	\label{app_eq:unconstrained}
	\bm{y}^* = \argmax_{\bm{y} \in \mathbb{R}^{p - \rank(\bm{A})}} f(\mathcal{B}_{\bm{A}} \bm{y})
	\end{equation}
	
	The two problems are equivalent in that characterizing the solutions to Equation~\ref{app_eq:constrained} fully characterizes the solutions to Equation~\ref{app_eq:unconstrained}, and vice versa. \citet[ch. 20]{lawson1974linear} discusses this in the case of least squares, but their discussion can be immediately generalized.
\end{enumerate}

\subsection{Proof of Theorem 1}

As noted in Theorem~\ref{thm:proper_prior}, the structured sparse prior has the following kernel where $\bar{\bm{D}}$ stacks together $\bm{D}$ and $\bm{F}_\ell$ vertically and $\bm{\beta} \in \mathbb{R}^{p}$.

\begin{equation}
\label{app:eq_priorkernel}
k(\bm{\beta}) = \exp\left(-\lambda\left[||\bm{D}\bm{\beta}||_1 + \sum_{\ell=1}^{L} \sqrt{\bm{\beta}^T\bm{F}_\ell\bm{\beta}}\right]\right)
\end{equation}

The prior is proper when the integral of $k(\bm{\beta})$ is finite for $\lambda > 0$. This occurs if and only if $\rank(\bar{\bm{D}}) = p$. The proof proceeds as follows:

First, I transform Equation~\ref{app:eq_priorkernel} using a decomposition of $\bm{F}_\ell$ such that $\bm{F}_\ell = \tilde{\bm{Q}}_\ell \tilde{\bm{Q}}_\ell^T$. An eigen-decomposition provides a natural choice (R2). Thus, $\sqrt{\bm{\beta}^T\bm{F}_\ell\bm{\beta}} = || \tilde{\bm{Q}}_\ell^T\bm{\beta}||_2$. Using the bounds on the $\ell_2$ norm by the $\ell_1$ norm (R1), the prior kernel can be bounded.

\begin{subequations}
	\begin{alignat}{2}
	&\exp\left(-\lambda\left[||\bm{D}\bm{\beta}||_1 + \sum_{\ell=1}^{L} || \tilde{\bm{Q}}^T_\ell \bm{\beta}||_1 \right]\right) \leq k(\bm{\beta}) \leq \exp\left(-\lambda\left[||\bm{D}\bm{\beta}||_1 + \frac{1}{\sqrt{p}} \sum_{\ell=1}^{L} || \tilde{\bm{Q}}^T_\ell \bm{\beta}||_1 \right]\right) \\
	\begin{split}&\exp\left(-\lambda ||\check{\bm{D}} \bm{\beta} ||_1\right) \leq k(\bm{\beta}) \leq \exp\left(-\lambda ||\bm{W}\check{\bm{D}} \bm{\beta} ||_1\right) \\ &\check{\bm{D}} = \left[\bm{D}^T, \tilde{\bm{Q}}_\ell, \cdots \right]^T \quad \bm{W} = \mathrm{bdiag}\left(\bm{I}_{K}, \frac{1}{\sqrt{p}} \bm{I}_{L \times p}\right)\end{split}
	\end{alignat}
\end{subequations}

Thus, examining the behavior of a structured sparse prior with only a penalty of $\check{\bm{D}}$ is sufficient to understand the behavior of a prior with $L > 0$ as $\bm{W}$ is always full rank. 

Consider the integral of the lower bound in two cases: First, assume that $\rank\left(\check{\bm{D}}\right) = p$. In this case, the integral is finite by upper-bounding (R1) and noting that $\check{\bm{D}}^T\check{\bm{D}}$ is full rank such that the integral is finite as an (invertible) change of variables could be applied by eigen-decomposing $\check{\bm{D}}^T\check{\bm{D}}$.

\begin{equation}
\int \exp\left(-\lambda ||\check{\bm{D}} \bm{\beta} ||_1\right) d\bm{\beta} \leq \int \exp\left(-\lambda ||\check{\bm{D}} \bm{\beta} ||_2\right) d\bm{\beta} = \int \exp\left(-\lambda \sqrt{\bm{\beta}^T \check{\bm{D}}^T\check{\bm{D}}\bm{\beta}}\right) d\bm{\beta} < \infty
\end{equation}

Thus, $\rank(\check{\bm{D}}) = p$ is sufficient for posterior propriety as $\bm{W} \check{\bm{D}}$ has the same rank as $\check{\bm{D}}$. Necessity can be proved in a similar way; assume that $\rank(\check{\bm{D}}) \neq p$. In this case, the following integral is infinite.

\begin{subequations}
	\begin{alignat}{2}
	\int \exp\left(-\lambda \normone{\check{\bm{D}} \bm{\beta}}_1\right) d\bm{\beta} &= \int \exp\left(-\lambda \normone{\check{\bm{U}}\left(\begin{array}{ll} \check{\bm{\Sigma}} & \bm{0} \\ \bm{0} & \bm{0} \end{array}\right) \bm{\theta} }_1\right) d\bm{\theta}  \\
	&= \int_{\mathbb{R}^{p - \rank(\check{\bm{D})})}} \left[\int_{\mathbb{R}^{\rank(\check{\bm{D}})}} \exp\left(-\lambda ||\check{\bm{U}}_{1} \check{\bm{\Sigma}} \bm{\theta}_{\mathcal{C}} ||_1\right) d\bm{\theta}_{\mathcal{C}}\right] d\bm{\theta}_{\mathcal{N}} = \infty
	\end{alignat}
\end{subequations}

The key move is to rotate $\bm{\beta}$ via an (invertible) transformation by multiplication such that $\bm{\theta} = \check{\bm{V}}^T\bm{\beta}$ where $\check{\bm{V}}$ is the right singular matrix coming from an SVD of $\check{\bm{D}}$ and $\check{\bm{U}}$ and $\check{\bm{\Sigma}}$ defined as in R2. For notation, denote $\bm{\theta}_\mathcal{C}$ as the first $\rank(\check{\bm{D}})$ elements of $\bm{\theta}$ and $\bm{\theta}_{\mathcal{N}}$ as the remaining elements. Note that since $\check{\bm{D}}$ is not full rank, the dimensionality of $\bm{\theta}_{\mathcal{N}}$ is at least one. Thus, the integral diverges with respect to $\bm{\theta}_{\mathcal{N}}$ but is finite with respect to $\bm{\theta}_{\mathcal{C}}$ from the above discussion as $\check{\bm{U}}_1 \check{\bm{\Sigma}}$ is full column rank. Thus, $\rank\left(\check{\bm{D}}\right) = p$ is a necessary condition.

Taken together, $\rank(\check{\bm{D}}) = p$ is thus necessary and sufficient for posterior propriety. Because of the bounds above, this thus describes behavior for the case of $\bar{\bm{D}}$. In that case, the normalizing constant can be expressed as $\lambda^p c$ where $c$ is a finite constant that depends only on $\bar{\bm{D}}$.

Theorem~\ref{thm:proper_prior} restates this more cleanly in terms of $\bm{F}_\ell$. Note that since $\bm{F}_\ell$ and $\tilde{\bm{Q}}_\ell^T$ have the same nullspace and rank, $\bar{\bm{D}}$ and $\check{\bm{D}}$ do likewise, and thus Theorem~\ref{thm:proper_prior} follows in the form expressed in the main text.

\subsection{Proof of Theorem 2}
The theorem assumes an un-normalized posterior of the following form where the likelihood is assumed to be log-concave with respect to $\bm{\eta}$. 

\begin{equation}
k(\bm{\beta} | \bm{y}) = L(\bm{\eta} | \bm{y})\exp\left(-\lambda \left[||\bm{D}\bm{\beta}||_1 + \sum_{\ell=1}^L \sqrt{\bm{\beta}^T\bm{F}_\ell\bm{\beta}}\right]\right) \quad \bm{\eta} = \bm{X}\bm{\beta}
\end{equation}

The theorem can be established by a similar method to the proof of prior propriety; again note the posterior kernel can be upper and lower bounded as follows:

\begin{equation}
L(\bm{\eta} | \bm{y}) \exp\left(-\lambda ||\check{\bm{D}} \bm{\beta} ||_1\right) \leq k(\bm{\beta} | \bm{y}) \leq L(\bm{\eta} | \bm{y}) \exp\left(-\lambda ||\bm{W}\check{\bm{D}} \bm{\beta} ||_1\right)
\end{equation}

Thus, it suffices to examine whether the posterior with a structured sparse prior on only $\check{\bm{D}}$ and $L = 0$ is proper. I thus consider the following posterior kernel in the subsequent analysis

\begin{equation}
k'(\bm{\beta} | \bm{y}) = L(\bm{\eta} | \bm{y}) \exp\left(-\lambda || \check{\bm{D}} \bm{\beta}||_1\right)
\end{equation}

To begin, I perform the above transformation to orthogonally rotate $\bm{\beta}$ by the right singular matrix from an SVD of $\check{\bm{D}}$: $\bm{\theta} = \check{\bm{V}}^T\bm{\beta}$. As noted in the main text, since this transformation is invertible, posterior inference on $\bm{\beta}$ is equivalent to performing inference on $\bm{\theta}$. Thus, the posterior on $\bm{\theta}$ can be expressed as follows where $\check{\bm{X}}_1$ and $\check{\bm{X}}_2$ represent the first $\rank(\check{\bm{D}})$ and remaining $p - \rank(\check{\bm{D}})$ columns of the rotated design ($\bm{X}\check{\bm{V}}$).

\begin{equation}
\label{app:theorem_2_D_form}
p(\bm{\theta}| \bm{y}) \propto L(\bm{\nu}' | \bm{y}) \exp\left(-\lambda ||\check{\bm{U}}_1\check{\bm{\Sigma}}\bm{\theta}_\mathcal{C}||_1\right) \quad \bm{\nu}' = \check{\bm{X}}_1 \bm{\theta}_\mathcal{C} + \check{\bm{X}}_2\bm{\theta}_{\mathcal{N}} \quad \check{\bm{X}} = \bm{X} \check{\bm{V}}
\end{equation}

Establishing the propriety of the posterior in Equation~\ref{app:theorem_2_D_form} can be done using results from \citet{michalak2016proper}. The paper provides a number of critical results summarized below as the following lemma:

\begin{lemma}
	\label{lemma:michalak_morris}
	\citet{michalak2016proper}: Assume a likelihood $L(\bm{\eta} | \bm{y})$ such that $\bm{\eta} = \bm{X}\bm{\beta}$. Define an exponentiated norm bound (ENB) as follows: An ENB holds if constants $c_0, c_1 > 0$ exist such that\footnote{They further remark (p. 550) that ``the constants $c_0$ and $c_1$ can be chosen independently of the $L_p$ norm, $p \geq 1$, because of norm equivalence, where two norms $L_p$ and $L_q$ on $\mathbb{R}^r$ are said to be norm-equivalent if and only if there exist constants $0 < c_2, c_2$ such that $c_2 ||\bm{v}||_p \leq ||\bm{v}||_q \leq c_3 ||\bm{v}_p$ for any vector $\bm{v}$. While $c_0$ and $c_1$ cannot depend on $\bm{\eta}$, they can depend on any know values including $\bm{y}$, $\bm{X}$, $\cdots$.''}
	\begin{equation}
	L(\bm{\eta} | \bm{y}) \leq c_0 \exp\left(-c_1 ||\bm{\eta}||\right)
	\end{equation}
	The following results hold:
	\begin{enumerate}
		\item If the likelihood as a function of $\bm{\eta}$ is log-concave and the MLE of $\bm{\eta}$ exists and is unique (or more broadly if $\bm{\eta}$ has multiple MLEs, all MLEs lie in a bounded set), then the likelihood has an ENB as a function of $\bm{\eta}$. (p. 550; Theorem 6, p. 561)
		\item For fixed $\bm{y}$, assume a likelihood $L(\bm{\eta} | \bm{y})$ as defined above has an ENB as defined above. Assume that $\bm{X}$ is full column rank and the prior density on $\bm{\beta}$ is bounded, i.e. $p(\bm{\beta}) \leq M < \infty$. Then, the posterior distributions of $\bm{\beta}$ and $\bm{\eta}$ are proper and $\bm{\beta}$ and $\bm{\eta}$ have proper posterior moment generating functions. (Theorem 1; p. 553).
		\item If $\bm{\beta}$ is entirely or partially known, let $\bm{\beta} = [\bm{\beta}_1^T, \bm{\beta}_2^T]^T$ so that $\bm{\eta} - \bm{X}_2 \bm{\beta}_2 = \bm{X}_1 \bm{\beta}_1$ with $\bm{X} = [\bm{X}_1, \bm{X}_2]$ partitioned accordingly with $\bm{\beta}_2$ known and $\bm{\beta}_1$ with Lebesque measure. This model has already been addressed by [the point above]. Because posterior propriety holds for all fixed $\bm{\beta}_2$, it holds when $\bm{\beta}_2$ has a proper prior distribution. (Remark 9; p. 559).
		\item GLMs [generalized linear models] with natural links are log-concave. Thus, a likelihood function $L(\bm{\eta} | \bm{y})$ for a GLM with a natural link and a finite MLE has an ENB as a function of $\bm{\eta}$. More generally, given a GLMM [generalized linear mixed model] (or other model) with a log-concave likelihood, it has an ENB if the MLE of $\bm{\eta}$ exists. (p. 551)
	\end{enumerate}
\end{lemma}

This lemma is crucial to establishing posterior propriety. As it is phrased in a rather general way, some remarks are in order to make clear the relevance to this paper. First, for the models considered, I assume that we are focused on models where the likelihood is log-concave. This includes most generalized linear models with standard choices of link functions. In general, their results could be applied to more complex models, but I leverage Remarks (1) and (4) from Lemma~\ref{lemma:michalak_morris} to use the existence of the MLE of $\bm{\eta}$ (and corresponding existence and uniqueness) to derive simple, easily verifiable, sufficient conditions for posterior propriety. Structured sparsity could be applied to more general models, but this requires more work to establish clear conditions for assessing the existence of an ENB. 

Second, note that the results are stated in terms of the existence and uniqueness of the MLE on $\bm{\eta}$ \emph{not} $\bm{\beta}$. This is designed to deal with the case of a rank deficient $\bm{X}$; adapting an example from \citet[p. 552]{michalak2016proper}, imagine that $\bm{X}$ had two identical columns. The MLE of $\bm{\beta}$ is clearly not unique although the MLE of $\bm{\eta}$ could be---as any MLE leads to the same $\bm{X}\bm{\beta}$.

Returning to the structured sparse case, note that Equation~\ref{app:theorem_2_D_form} reflects the scenario described in Remark (3) of Lemma~\ref{lemma:michalak_morris} where $\bm{\theta}_{\mathcal{C}}$ has a proper prior and there is a flat prior on $\bm{\theta}_{\mathcal{N}}$. I prove the following Lemma:

\begin{lemma}
	\label{lemma:theorem_D}
	For the likelihood described in Equation~\ref{app:theorem_2_D_form}, define $\hat{\bm{\theta}}_\mathcal{N}$ as follows.
	\begin{equation}
	\hat{\bm{\theta}}_\mathcal{N} = \arg\max_{\bm{\gamma}} \ln L(\bm{\psi} | \bm{y}); \quad \bm{\psi} = \check{\bm{X}}_2 \bm{\gamma}
	\end{equation} 
	
	If $\hat{\bm{\theta}}_\mathcal{N}$ exists and is unique, then the posterior on $p(\bm{\theta} | \bm{y})$ is proper.
\end{lemma}

This can be proved by directly applying \citet{michalak2016proper}'s results summarized in Lemma~\ref{lemma:michalak_morris}. If the MLE on $\bm{\theta}_{\mathcal{N}}$ exists and is unique, then this ensures that the MLE on $\bm{\nu}'$ exists and is unique for $\bm{\theta}_{\mathcal{C}} = 0$ as it is for $\bm{\psi}$. Thus, Remarks (1) and (2) from Lemma~\ref{lemma:michalak_morris} apply and the posterior is proper for $\bm{\theta}_{\mathcal{C}} = \bm{0}$. 

Note, however, that for any choice of $\bm{\theta}_{\mathcal{C}} \in \mathbb{R}^{\rank(\check{\bm{D}})}$, the posterior remains proper. Since the MLE of $\bm{\nu}'$ exists and is unique when $\bm{\theta}_{\mathcal{C}} = \bm{0}$, it can be simply shifted to account for a non-zero offset.\footnote{Put another way, if the MLE exists and is unique, then any finite ``offset'' will still have an MLE that exists and is unique.} Thus, Remark (3) from Lemma~\ref{lemma:michalak_morris} applies as there is a proper prior on $\bm{\theta}_{\mathcal{C}}$ by Theorem~\ref{thm:proper_prior} and noting that $\check{\bm{U}}_1\check{\bm{\Sigma}}$ is full column rank. Thus, the posterior on $p(\bm{\theta} | \bm{y})$ is proper and thus a structured sparse prior on $\bm{\beta}$ with $\check{\bm{D}}$ is proper as $k'(\bm{\beta} | \bm{y})$ has a finite integral. As this (or a finite transformation) upper bounds the original posterior kernel $k(\bm{\beta} | \bm{y})$, this also ensures the original posterior on $\bm{\beta}$ is proper. 

Condition (b) in Theorem~\ref{thm:proper_posterior} expresses the claim in Lemma~\ref{lemma:theorem_D} slightly differently. It states that if $\hat{\bm{\beta}}_{\mathcal{N}(\bm{D})}$, as defined below, exists and is unique, then the posterior is proper.

\begin{equation}
\label{app:eq_restate_thm2b}
\hat{\bm{\beta}}_{\mathcal{N}(\bm{D})} = \argmax_{\bm{\beta}} L(\bm{\eta} | \bm{y}) \quad \mathrm{s.t.} \quad \bar{\bm{D}} \bm{\beta} = \bm{0}
\end{equation}

The equivalence between this condition and the one defined in Lemma~\ref{lemma:theorem_D} follows in three parts. First, note that $\hat{\bm{\theta}}_{\mathcal{N}}$ can be expressed as the optimization over the entire $\bm{\theta}$ space subject to a linear constraint that $\bm{\theta}_\mathcal{C} = \bm{0}$. Equation~\ref{app_eq:restate_restrict} restates the condition in Lemma~\ref{lemma:theorem_D}.

\begin{equation}
\label{app_eq:restate_restrict}
\hat{\bm{\theta}}= \arg\max_{\bm{\theta} \in \mathbb{R}^p} \ln L(\bm{\nu}' | \bm{y}); \quad \bm{\nu}' = \check{\bm{X}}_1 \bm{\theta}_\mathcal{C} + \check{\bm{X}}_2 \bm{\theta}_\mathcal{N} \quad \bm{\theta}_\mathcal{C} = \bm{0}
\end{equation}

Second, note that since $\bm{\theta} = \check{\bm{V}}^T\bm{\beta}$, the problem can be rotated back into the $\bm{\beta}$ space although the optimization is now over a linear subspace defined by the span of the columns of $\check{\bm{V}}_2$ as $\bm{\beta} = \check{\bm{V}}_1 \bm{\theta}_\mathcal{C} + \check{\bm{V}}_2 \bm{\theta}_{\mathcal{N}} = \check{\bm{V}}_2 \bm{\theta}_{\mathcal{N}}$ where $\bm{\theta}_\mathcal{N} \in \mathbb{R}^{\rank(\check{\bm{D}})}$. This, however, is exactly the nullspace of $\check{\bm{D}}$. By noting the equivalence of optimization over the nullspace and a model with a linear constraint (R4), Equation~\ref{app_eq:nullcheck_D} follows. Since $\check{\bm{D}}$ and $\bar{\bm{D}}$ have equivalent nullspaces, the phrasing in Theorem~\ref{thm:proper_posterior} follows.

\begin{equation}
\label{app_eq:nullcheck_D}
\hat{\bm{\beta}}= \arg\max_{\bm{\beta}} \ln L(\bm{\nu} | \bm{y}); \quad \bm{\nu} = \bm{X}\bm{\beta} \quad \mathrm{s.t.} \quad \check{\bm{D}} \bm{\beta} = \bm{0}
\end{equation}

Finally, condition (a)---the necessary condition---follows easily by examining the rank of $\check{\bm{X}}_2$. This must be full rank for the posterior to be proper; the proof proceeds by contradiction: Assume that $\rank(\check{\bm{X}}_2) < p - \rank(\bm{D})$, i.e. it was not full column rank. By the same logic that Theorem~\ref{thm:proper_prior} is shown to be necessary, it is clear that the integral of the kernel in Equation~\ref{app:theorem_2_D_form} diverges in this case as after an orthogonal rotation of $\bm{\theta}_\mathcal{N}$ by the singular value decomposition of $\check{\bm{X}}_2$. This is because after such a rotation, there are some elements of the rotated $\bm{\theta}_{\mathcal{N}}$ that appear nowhere in the prior nor the likelihood.

Recall that $\check{\bm{X}}_2$ equals $\bm{X}$ times the basis for a nullspace of $\check{\bm{D}}$. The condition that it is full rank is thus equivalent to ensuring that the nullspaces of $\check{\bm{D}}$ and $\bm{X}$ intersect only at $\bm{0}$ or that their stacked matrix is full rank (R3). This is thus the condition as stated in Theorem~\ref{thm:proper_posterior} again noting that $\check{\bm{D}}$ and $\bar{\bm{D}}$ have the same nullspace.

It can also be derived directly by rotating $\bm{\beta}$ by the right singular matrix coming from the stacked SVD of $\bm{X}$ and $\bar{\bm{D}}$. In that case, if it is not full rank, there exist some components in the rotated space that no longer appear in the posterior and thus the integral over those components diverges.

\subsubsection{Proof of Corollary 1}

Corollary~\ref{coro:glm} can be proven straightforwardly. For the linear model, \citet[ch. 20]{lawson1974linear} contains the basic idea. Assume Condition (a) holds. This implies that $\check{\bm{X}}_{2}$ is full rank. For a linear model, that ensures a single unique MLE and thus (a) implies (b). As (b) is sufficient for propriety, (a) is necessary and sufficient for posterior propriety. (a) and (b) is thus a slightly redundant way of stating this claim.

For the multinomial case with a standard link (logit or probit), results in \citet{speckman2009multinomial} can be employed. Specifically, their Theorem 3 restated as a lemma notes:

\begin{lemma}
	\label{lemma:speckman}
	\citet[p. 742]{speckman2009multinomial}: For the multinomial logistic or probit choice model, the following
	conditions are equivalent.
	\begin{enumerate}
		\item There is overlap in the sample [see paper for discussion]
		\item The MLE of $\bm{\theta}$ exists and is finite.
		\item The posterior of $\bm{\theta}$ is proper under the constant prior.
	\end{enumerate}
\end{lemma}

For Corollary~\ref{coro:glm}, assume that Conditions (a) and (b) hold. This implies that $\check{\bm{X}}_2$ is full rank. Thus, if the MLE of $\hat{\bm{\theta}}_{\mathcal{N}}$ is finite (exists), it is unique. This implies that if Condition (b) is satisfied, Lemma~\ref{lemma:speckman} immediately applies to ensure the posterior of $p(\bm{\theta}_\mathcal{N} | \bm{y})$ is proper if $\bm{\theta}_\mathcal{C}$ is fixed. By the logic above, since there is a proper prior on $\bm{\theta}_{\mathcal{C}}$, the entire posterior on $p(\bm{\theta} | \bm{y})$ is proper and thus the posterior on $\bm{\beta}$ is proper.

Thus, since (a) and (b) are jointly sufficient, they are jointly necessary and sufficient as (a) alone is necessary.

\subsection{Proof of Theorems 3 and 4}

Theorem~\ref{thm:gibbs} can be established by extending results (e.g. \cite{park2008bayesian,kyung2010penalized}). They note the following identity:

\begin{subequations}
	\label{app_eq:andrews}
	\begin{alignat}{2}
	\int_0^\infty \frac{1}{\sqrt{2\pi \tau^2}} \exp\left(\frac{-z^2}{2\tau^2} - \frac{\lambda^2 \tau^2}{2}\right) \frac{\lambda^2}{2}d\tau^2 = \frac{\lambda}{2}\exp(-\lambda|z|) \\
	z \sim N(0, \tau^2); \quad \tau^2 \sim \mathrm{Exp}(\lambda^2/2) 
	\end{alignat}
\end{subequations}

The first line is crucial for our purposes as it provides a way to rewrite each $K$ and $L$ penalty; Equation~\ref{app_eq:andrews} applies it to each term to get the joint density in Theorem~\ref{thm:gibbs}. Assume a proper structured sparse prior with normalizing constant $\lambda^p c$.

\begin{align}
\label{app:eq_joint}
&p\left(\bm{\beta}, \{\tau^2_k\}_{k=1}^K, \{\xi^2_\ell\}_{\ell=1}^L | \lambda \right) = \lambda^p c~\times &\begin{split} &\prod_{k=1}^K \frac{2}{\lambda} \cdot \frac{1}{\sqrt{2\pi\tau^2_k}} \exp\left(-\frac{\bm{\beta}^T\bm{d}_k \bm{d}_k^T\bm{\beta}}{2\tau^2_k} - \frac{\lambda^2\tau^2_k}{2}\right) \cdot \lambda^2/2~\times \\
&\prod_{\ell=1}^{L} \frac{2}{\lambda} \cdot \frac{1}{\sqrt{2\pi\xi^2_\ell}}\exp\left(-\frac{\bm{\beta}^T\bm{F}_\ell \bm{\beta}}{2\xi_\ell^2} - \frac{\xi_\ell^2\lambda^2}{2}\right) \cdot \lambda^2/2
\end{split}
\end{align}

Ignoring constants that do not depend on the parameters gives the result in Theorem~\ref{thm:gibbs}. Note that the marginal prior this implies on $\{\tau^2_k\}$ and $\{\xi^2_\ell\}$ will not be the simple independent product of Gamma random variables outside of very special choices of $\bm{D}$ and $\bm{F}_\ell$ that are used in prior research (e.g. \cite{park2008bayesian,kyung2010penalized}) and thus working from the joint prior is required to sample $\bm{\beta}$. 

The Gibbs Sampler follows by a change of variables. Consider a single $\tau^2_k$. The full conditional is proportional to the following; applying a change of variables gives a density that is Inverse Gaussian. The density of the Inverse Gaussian comes from \citet{park2008bayesian} where $\mu, \lambda > 0$.

\begin{subequations}
	\begin{alignat}{2}
	p(\tau^2_k | -) &\propto (\tau^2_k)^{-1/2}\exp\left(-\frac{\bm{\beta}^T\bm{d}_k \bm{d}_k^T\bm{\beta}}{2\tau^2_k} - \frac{\lambda^2\tau^2_k}{2}\right) \\
	p(1/\tau^2_k | -) &\propto (\tau^2_k)^{-3/2} \exp\left(-\lambda^2/2 (1/\tau^2_k)^{-1} - \frac{1}{2} \left[\bm{\beta}^T\bm{d}_k\right]^2 \cdot (1/\tau^2_k)\right) \\
	1/\tau^2_k &\sim \mathrm{InvGaussian}\left(\frac{\lambda}{|\bm{d}_k^T\bm{\beta}|}, \lambda^2\right) \\
	x &\sim \mathrm{InvGaussian}(\mu, \lambda) \quad \mathrm{iff} \quad p(x | \mu, \lambda) =  \sqrt{\frac{\lambda}{2\pi}} x^{-3/2} \exp\left(-\frac{\lambda ( x - \mu)^2}{2\mu^2 x}\right)
	\end{alignat}
\end{subequations}

Thus, the full conditionals on all of the augmentation variables $\{\tau^2_k\}$ and $\{\xi^2_\ell\}$ are conditional independent given $\bm{\beta}$ (and $\lambda$) and all have Inverse Gaussian densities as implied by the above equation and stated in Theorem~\ref{thm:gibbs}.

Finally, note that $\lambda$ can be sampled as well. As is common in this literature, a Gamma prior is placed on $\lambda^2$ as it is conditionally conjugate with the joint density (\cite{park2008bayesian,kyung2010penalized}). For some proper Gamma prior $(a_0, b_0)$ (shape-rate parameterization), the full conditional on $\lambda^2$ is shown below

\begin{equation}
\lambda^2 | - \sim \mathrm{Gamma}\left(a_0 + \frac{p + K + L}{2}, b_0 + \frac{1}{2} \left[\sum_{k=1}^K \tau^2_k + \sum_{\ell=1}^{L}\xi^2_\ell\right] \right)
\end{equation}

\subsection{Extension to Global-Local Priors}
\label{app:sub_global_local}

The above discussion used a particular form of penalization ($\ell_1$ and $\ell_2$ norms) to create sparse estimates. A large literature has developed alternative Bayesian methods based on mixtures of normal distributions known as global-local priors (\cite{polson2011shrink}). This includes popular methods such as the adaptive LASSO, horseshoe, and others. Theorem~\ref{thm:global_local} shows that the above results are not specific to LASSO-type penalties.

\begin{theorem}{Application to Global-Local Priors}
	\label{thm:global_local}
	
	Assume that the prior on $\delta$ is a global-local prior \citep{polson2011shrink} whose marginal density $p_{g, \bm{\lambda}}(\delta)$ can be expressed as follows, where $\bm{\lambda}$ is a fixed vector of hyper-parameters and $g$ is some proper probability distribution whose support is on the non-negative reals:
	
	$$p_{g,\bm{\lambda}}(\delta) = \int_0^\infty (2\pi\tau^2)^{-1/2} \exp\left(-\frac{\delta^2}{2\tau^2}\right) g(\tau^2; \bm{\lambda}) d\tau^2$$
	
	First, define the following generalization to a multivariate $\bm{\delta}$ as follows:
	
	$$p_{g, \bm{\lambda}}(\bm{\delta}) = \int_0^\infty (2\pi\tau^2)^{-p/2}\exp\left(-\frac{\bm{\delta}^T\bm{\delta}}{2\tau^2}\right) g(\tau^2; \bm{\lambda}) d\tau^2$$
	
	Further, define a global-local structured sparse prior as having the following density:
	\begin{alignat}{2}
	p\left(\bm{\beta}, \{\tau^2_k\}_{k=1}^K, \{\xi^2_\ell\}_{\ell=1}^L | \bm{\lambda}\right) \propto \begin{split}
	&\exp\left(-\frac{1}{2}\bm{\beta}^T\left[\sum_{k=1}^K \frac{\bm{d}_k\bm{d}_k^T}{\tau^2_k} + \sum_{\ell=1}^L \frac{\bm{F}_\ell}{\xi^2_\ell}\right]\bm{\beta}\right) \times \\
	&\prod_{k=1}^K \frac{g(\tau^2_k; \bm{\lambda})}{(\tau^2_k)^{1/2}} \prod_{\ell=1}^L \frac{g(\xi^2_\ell; \bm{\lambda})}{(\xi^2_\ell)^{p/2}}
	\end{split} 
	\end{alignat}
	
	Theorems~\ref{thm:proper_prior} and~\ref{thm:proper_posterior} characterize the marginal prior on $\bm{\beta}$.
\end{theorem}

The proof can be established in two parts. First, I generalize Equation~\ref{app_eq:global_local_multi} to the case of a positive semi-definite $\bm{F}$ matrix, noting that if it is not positive definite, the prior is improper over $\bm{\delta}$. 

\begin{equation}
\label{app_eq:global_local_multi}
p_{g, \bm{\lambda}}(\bm{\delta}) = \int_0^\infty (2\pi\tau^2)^{-p/2}\exp\left(-\frac{\bm{\delta}^T\bm{\delta}}{2\tau^2}\right) g(\tau^2; \bm{\lambda}) d\tau^2 \quad \bm{\delta} | \tau^2 \sim N(\bm{0}, \bm{I}_p \cdot \tau^2)~\tau^2 \sim g(\tau^2;\bm{\lambda})
\end{equation}

For notational clarity, if I suppress $\bm{F}$ from the notation it is assumed to be the identity matrix. The product of this kernel and the analogous kernel for a linear restriction gives the proposed joint density in Theorem~\ref{thm:gibbs}.

\begin{equation}
p_{g, \bm{\lambda}, \bm{F}}(\bm{\delta}) = \int_0^\infty (2\pi\tau^2)^{-p/2}\exp\left(-\frac{\bm{\delta}^T\bm{F}\bm{\delta}}{2\tau^2}\right) g(\tau^2; \bm{\lambda}) d\tau^2
\end{equation}

With this in hand, the marginal structured sparse global-local prior on $\bm{\beta}$ has the following kernel if one integrates away the $\{\tau^2_k\}$ and $\{\xi^2_\ell\}$. Equation~\ref{app:global_local_marg}b follows by noting that since $\bm{F}_\ell$ is positive semi-definite, it can be replaced with an identity matrix.

\begin{subequations}
	\label{app:global_local_marg}
	\begin{alignat}{2}
	k_{g,\bm{\lambda}}(\bm{\beta}) &= \prod_{k=1}^K p_{g,\bm{\lambda}}\left(\bm{d}_k^T\bm{\beta}\right) \cdot \prod_{\ell=1}^{L}p_{g,\bm{\lambda}, \bm{F}}\left(\bm{\beta}\right) \\
	&= \prod_{k=1}^K p_{g,\bm{\lambda}}\left(\bm{d}_k^T\bm{\beta}\right) \cdot \prod_{\ell=1}^{L}p_{g,\bm{\lambda}}\left(\tilde{\bm{Q}}^T_\ell\bm{\beta}\right)  \\
	& = f(\check{\bm{D}} \bm{\beta})
	\end{alignat}
\end{subequations}

Consider Theorem~\ref{thm:proper_prior}: Using the notation in the earlier proofs, assume that $\rank\left(\check{\bm{D}}\right) \neq p$. The prior is improper by the same logic as before; if one defines $\bm{\theta} = \check{\bm{V}}^T \bm{\beta}$, i.e. the orthogonal rotation by the right singular matrix from a SVD of $\check{\bm{D}}$, there again are some components that appear nowhere in the prior. Thus, the integral of the kernel in Equation~\ref{app:global_local_marg} diverges.

Sufficiency must be proven differently from above as the $\ell_1$ and $\ell_2$ bounds are no longer permissible. It can be shown in the following way: Assume $\check{\bm{D}}$ has full column rank. It can be expressed as follows $\check{\bm{D}} = \check{\bm{U}}_1 \bm{\Sigma} \check{\bm{V}}^T$ where the full column rank means that even for the ``thin'' SVD, $\check{\bm{V}}^T$ is invertible. As before, I thus orthogonally rotate $\bm{\beta}$ to define $\bm{\theta}' = \bm{\Sigma} \check{\bm{V}}^T\bm{\beta}$. 

\begin{subequations}
	\label{app_eq:colspace}
	\begin{alignat}{2}
	\int_{\mathbb{R}^p} f(\check{\bm{D}}\bm{\beta}) d\bm{\beta} &= \det(\bm{\Sigma})^{-1} \int_{\mathbb{R}^p} f(\check{\bm{U}}_1 \bm{\theta}') d\bm{\theta}' \\
	&=  \det(\bm{\Sigma})^{-1} \int_{\bm{z} \in \mathcal{C}(\check{\bm{D}})} f(\bm{z}) d\bm{z} \\
	&\leq \det(\bm{\Sigma})^{-1} \int_{\mathbb{R}^{K+ p \cdot L}} f(\bm{z}) d\bm{z} < \infty
	\end{alignat}
\end{subequations}

This manipulation moves from the left to the right-hand side of Equation~\ref{app_eq:colspace}a. The following line notes that $\check{\bm{U}}_1 \bm{\theta}'$ is a vector of length $\mathbb{R}^{K + p \times L}$. Given that $\check{\bm{U}}_1$ is a basis for the column space of $\check{\bm{D}}$ (denoted by $\mathcal{C}(\check{\bm{D}})$), this is thus equivalent to an integral over a particular subspace of $\mathbb{R}^{K + p \times L}$. 

Equation~\ref{app_eq:colspace}c follows by noting that since $f(\bm{z})$ is non-negative since it is the product of probability density functions, the integral in column space must be weakly smaller than the integral over the entire $\mathbb{R}^{K + p \times L}$ space. That integral over the entire space, however, is simply the product of (proper) probability density functions and thus is finite.

Thus, as in the LASSO case, $\check{\bm{D}}$ being full rank is necessary and sufficient for posterior propriety. Thus, by the same logic as before, examining the rank of $\bar{\bm{D}}$ characterizes the prior propriety of a global-local structured sparse prior. Note, however, that the claim following Theorem~\ref{thm:proper_prior} about the characterization of the normalizing constant does not necessarily apply.

Equation~\ref{app_eq:colspace}b has another interesting interpretation. One can think of a structured sparse prior as ``similar'' to the product-of-independent sparsity inducing priors \emph{but} that the values are constrained to lie in the column space of $\check{\bm{D}}$ to deal with the linear constraints imposed by the fact that the elements of $\bm{z}$ are not allowed to freely vary.

The final point is to show that the Theorem~\ref{thm:proper_posterior} applies. Fortunately, nothing in that proof was specific to the $\ell_1/\ell_2$ bounds except using the results of Theorem~\ref{thm:proper_prior} and thus it follows automatically.

\section{Calibrating $\lambda$}
\label{app:calibrate_lambda}

There is a question of how to choose the strength of the prior ($\lambda$) for both the Bayesian and non-Bayesian approaches. This can be done in a variety of ways; cross-validation is a popular option but requires fitting the model repeatedly and thus may be computationally expensive. It further requires a ``simple'' data structure that can be easily partitioned into separate folds. 

An alternative strategy uses information criterion such as the AIC or BIC. That requires evaluating the log-likelihood as well as a measure of complexity of the model. \citet{tibshirani2012df} provide an unbiased measure of the degrees of freedom for the generalized LASSO (arbitrary $\bm{D}$; $L = 0$) in the linear model that can be used to calculate these information criteria. For the non-linear case, a common approach is to use the same criterion.

In the fully Bayesian setting, one often prefers to set a prior on $\lambda$ and sample it alongside $\bm{\beta}$. Appendix~\ref{app:proof_ssp} shows that this can be easily done with a standard conditionally conjugate prior on $\lambda^2$ (\cite{park2008bayesian}). Calibrating the prior, however, raises similar questions to the non-Bayesian case. 

Section~\ref{section:est_het} uses a hybrid strategy where the fast posterior mode algorithm and AIC is used to find a plausible $\lambda$ to anchor the prior. Specifically, I use the fast EM algorithm to perform a grid search over $\lambda$ and choose the model with the best AIC. Given the optimal $\lambda^*$ from this grid search, I place a conditionally conjugate prior on $\lambda^2$ (see Appendix~\ref{app:proof_ssp}) where the mean and median are $(\lambda^*)^2$. As I show, the results are very similar if $\lambda$ is frozen at $\lambda^*$ in the Bayesian analysis.

\section{Details of Inference}
\label{app:EM_multinomial}

This section derives the Gibbs Sampler algorithms for the linear and multinomial models. It then discusses particularities of the EM algorithm.
\subsection{Linear Regression}

For linear regression, we need to incorporate the error variance $\sigma^2$ into the model. Assume the following generative framework following \citet{park2008bayesian} where $\rank(\bar{\bm{D}}) = m$:

\begin{subequations}
	\begin{alignat}{2}
	\bm{y} | \bm{\beta}, \sigma^2 &\sim N\left(\bm{X}\bm{\beta}, \sigma^2\bm{I}_{N}\right) \\
	p(\bm{\beta} | \lambda^2, \sigma^2) &\propto w_{\bar{\bm{D}}} \lambda^m/\sigma^m \exp\left(-\frac{\lambda}{\sigma} \left[||\bm{D}\bm{\beta}||_1 + \sum_{\ell=1}^{L} \sqrt{\bm{\beta}^T\bm{F}_\ell\bm{\beta}}\right]\right) 
	\end{alignat}
\end{subequations}

The log-posterior, including a prior of $p_0(\sigma^2)$ on $\sigma^2$ and $p_0(\lambda^2)$ on $\lambda^2$ can be written as follows, up to constant involving $\bm{D}$:

\begin{alignat}{2}
\begin{split}
\ln p(\bm{\beta} | \lambda^2, \sigma^2) &\propto -\frac{N}{2} \ln (2\pi\sigma^2) - \frac{||\bm{y} - \bm{X}\bm{\beta}||_2^2}{2\sigma^2} + \\ &m \ln(\lambda) - m/2 \ln(\sigma^2) - \frac{\lambda}{\sigma} \left[||\bm{D}\bm{\beta}||_1 + \sum_{\ell=1}^{L} \sqrt{\bm{\beta}^T\bm{F}_\ell\bm{\beta}}\right] + \ln p_0(\sigma^2) + \ln p_0(\lambda^2)
\end{split}
\end{alignat}

The joint prior on $\bm{\beta}, \sigma^2, \{\tau^2_k\}, \{\xi^2_\ell\}$ follows:

\begin{equation}
\begin{split}
p(\bm{\beta}, \sigma^2, \{\tau^2_k\}, \{\xi^2_\ell\}| \lambda) \propto  &\exp\left(-\frac{1}{2\sigma^2} \cdot \bm{\beta}^T\left[\sum_{k=1}^K \frac{\bm{d}_k\bm{d}_k^T}{\tau^2_k} + \sum_{\ell=1}^{L} \frac{\bm{F}_\ell}{\xi^2_\ell}\right]\bm{\beta}\right) \times \\
&(\sigma^2)^{-m/2} p_0(\sigma^2) \cdot \lambda^{K+L+m} \prod_{k=1}^K \frac{\exp(-\lambda^2/2 \cdot \tau^2_k)}{\sqrt{\tau^2_k}} \prod_{\ell=1}^{L} \frac{\exp(-\lambda^2/2 \cdot \xi^2_\ell)}{\sqrt{\xi^2_\ell}}
\end{split}
\end{equation}

From this, the full conditional for $\sigma^2$ becomes, assuming a conjugate prior of $p_0(\sigma^2) \sim \mathrm{InverseGamma}(a_0, b_0)$, $p_0(\lambda^2) \sim \mathrm{Gamma}(a_{0,\bm{\Lambda}}, b_{0, \bm{\Lambda}})$ and $m = rank(\bar{\bm{D}})$.

\begin{alignat}{2}
\sigma^2 | - \sim \mathrm{InverseGamma}\left(a_{0,\sigma} + \frac{1}{2}\left[N + m\right], b_{0,\sigma} + \frac{1}{2}\left[
\begin{aligned}(\bm{y}-  \bm{X}\bm{\beta})^T(\bm{y}-\bm{X}\bm{\beta}) + \\ \bm{\beta}^T\left[\sum_{k=1}^{K}\frac{\bm{d}_k\bm{d}_k^T}{\tau^2_k} + \sum_{\ell=1}^{L} \frac{\bm{F}_\ell}{\xi^2_\ell}\right]\bm{\beta}\end{aligned}\right]\right)
\end{alignat}

The full conditionals on the other parameters are easily derived.

\begin{subequations}
	\begin{alignat}{3}
	\bm{\beta} | - &\sim N\left(\bm{\Sigma}_\beta\bm{X}^T\bm{y}, \sigma^2 \bm{\Sigma}_\beta\right); \quad \bm{\Sigma}_\beta = \left[\bm{X}^T\bm{X} + \sum_{k=1}^{K}\frac{\bm{d}_k\bm{d}_k^T}{\tau^2_k} + \sum_{\ell=1}^{L} \frac{\bm{F}_\ell}{\xi^2_\ell}\right]^{-1}\\
	1/\tau^2_k | - &\sim \mathrm{InvGaussian}\left(\frac{\lambda \sigma}{|\bm{d}_k^T\bm{\beta}|}, \lambda^2\right) \quad 1/\xi^2_\ell | -\sim \mathrm{InvGaussian}\left(\frac{\lambda \sigma}{\sqrt{\bm{\beta}^T\bm{F}_\ell\bm{\beta}}}, \lambda^2\right) \\
	\lambda^2 | - &\sim \Gm\left(a_{0, \Lambda} + [K+L+m]/2, \quad b_{0, \Lambda} + \frac{1}{2}\sum_{k=1}^K \tau^2_k + \frac{1}{2} \sum_{\ell=1}^{L} \xi^2_\ell \right)
	\end{alignat}
\end{subequations}

\subsection{Multinomial Regression}

Inference is derived for a $C$-category multinomial regression with the logistic regression being a special case. Denote the observation as $y_i$ as taking on values from $1$ to $C$. For simplicity, I assume the covariates are equal across levels. For each $y_{i}$, the generative model is multinomial:

\begin{equation}
p(y_{i} = c | \{\bm{\beta}_{c}\})  \propto \exp(\bm{x}_{i}^T\bm{\beta}_{c})
\end{equation}

The likelihood is shown below, setting $\bm{\beta}_C = \bm{0}$ to identify the model.

\begin{equation}
\prod_{i=1}^N \left[\frac{\exp(\bm{x}_{i}^T\bm{\beta}_{c})}{\sum_{l=1}^C \exp(\bm{x}_{i}^T\bm{\beta}_{l})}\right]^{I(y_{i} = c)}
\end{equation}

Structured sparsity, as before, can be encoded by placing priors on $\bm{\beta}_c$. I focus on the case of identical structures for each $\bm{\beta}_c$, but one could impose more complex restrictions by constraining coefficients across-levels $c$ in theory. The prior has the following form:

\begin{equation}
p(\{\bm{\beta}_c\}) \propto \prod_{c=1}^{C-1} \lambda^m \exp\left(-\lambda \left[||\bm{D}\bm{\beta}_c|| + \sum_{\ell=1}^{L} \sqrt{\bm{\beta}^T\bm{F}_\ell\bm{\beta}}\right]\right)
\end{equation}

A Gibbs Sampler can be constructed following \citet{polson2013polyagamma}. The key idea is to cycle through $c$ and perform inference conditional on all other $\bm{\beta}_c$:

\begin{equation}
\begin{split}
&p(\{\bm{\beta}_{c}\} | \{\bm{\beta}_{\neg c}\}) = \prod_{i=1}^N \frac{\exp(\bm{x}_{i}^T\bm{\beta}_{c} - O_{ic})^{I(y_{i} = c)}}{\exp(\bm{x}_{i}^T\bm{\beta}_{c} - O_{ic}) + 1} \cdot p(\bm{\beta}_c); O_{ic} = \ln\left(\sum_{l \neq c} \exp(\bm{x}_{i}^T\bm{\beta}_{l})\right)
\end{split}
\end{equation}

For each $c$, one can perform Polya-Gamma augmentation as outlined in \citet{polson2013polyagamma}. The core identity is that, for $\omega \sim PG(1,x)$ where $PG$ is a Polya-Gamma random variable---a particular infinite convolution of Gamman random variables:\footnote{Specifically, a Polya-Gamma variable is defined as below; see \citet{polson2013polyagamma} for details.
	
	$$\omega = \frac{1}{2\pi^2} \sum_{i=1}^\infty \frac{Z_i}{(k-1/2)^2 + c^2/(4\pi^2)}; \quad Z_i \sim^{i.i.d.} Gamma(b,1)$$
}  

\begin{subequations}
	\begin{alignat}{3}
	\bm{\beta}_{c} | \{\bm{\beta}_{\neg c}\} &\propto \prod_{i=1}^N \frac{\exp(\bm{x}_{i}^T\bm{\beta}_{c} - C_{ic})^{I(y_{i} = c)}}{\exp(\bm{x}_{i}^T\bm{\beta}_{c} - O_{ic}) + 1} \times p(\bm{\beta}_c) \\
	\omega_{i,c} | \bm{\beta}_c, \{\bm{\beta}_{\neg c}\} &\sim PG(1, \bm{x}_i^T\bm{\beta}_c - O_{ic}) \\
	\begin{split}
	\bm{\beta}_c | \{\omega_{i,c}\}, \{\bm{\beta}_{\neg c}\} \sim &N\left(\bm{\Lambda}^{-1}_\beta \bm{X}^T\bm{s}, \quad \bm{\Lambda}_\beta^{-1}\right) \quad \bm{\Lambda}_\beta = \left[\sum_i \omega_{i,c} \bm{x}_i\bm{x}_i^T \right]\\
	&s_i = I(y_i = c) - 1/2 - \omega_i (\bm{x}_i^T\bm{\beta}_c - O_{ic}); \quad [\bm{s}]_i = s_i
	\end{split}
	\end{alignat}
\end{subequations}

This manipulation occurs independently of the data augmentation for the sparsity penalty. Thus, one can sample the $\{\tau^2_k\}$ as before and thus create a posterior on $\bm{\beta}_c$ as follows

\begin{equation}
\begin{split}
&\bm{\beta}_c | \{\omega_{i,c}\}, \{\bm{\beta}_{\neg c}\}, \{\tau^2_k\}, \{\xi^2_\ell\} \sim N\left(\bm{\Sigma}_\beta \bm{X}^T \bm{s}, \bm{\Sigma}_\beta\right); \bm{\Sigma}_\beta = \left[\bm{\Lambda}_\beta + \sum_{k=1}^K \frac{\bm{d}_k\bm{d}_k^T}{\tau^2_k} + \sum_{\ell=1}^{L} \frac{\bm{F}_\ell}{\xi^2_\ell}\right]^{-1}
\end{split}
\end{equation}

\subsection{EM Algorithm}

The EM algorithm follows automatically from the above results. The $E$-Step ($\tau^2_k; \omega_{i,c}$) is tractable and the $M$-Step (expectation of log complete posterior) is a simple ridge regression. It iterates these until convergence (e.g. stationarity of log-posterior or parameters).

There is one subtle point: If $\bm{d}_k^T\bm{\beta} = 0$ or $\bm{\beta}^T\bm{F} \bm{\beta} = 0$, then the corresponding augmentation variable $\tau^2_k$ or $\xi^2_\ell$ no longer has a proper density. Thus, one must deal with the fact that as $\bm{d}_k^T\bm{\beta} \to 0$, $E[1/\tau^2_k] \to \infty$ which may cause numerical instability in the algorithm. \citet{polson2011svm} suggest that when a restriction nearly binds (e.g. $|\bm{d}_k^T\bm{\beta}| < 10^{-6}$), it should be treated as binding for the remaining iterations, i.e. require that $\bm{d}_k^T\bm{\beta} = 0$ in each subsequent iteration.

I follow this logic but note their strategy relies on restricted least squares that cannot be applied to arbitrary structures. Thus, I adapt an older strategy from \citet{lawson1974linear} discussed above and perform (unrestricted) inference in the nullspace of binding restrictions and then back-out the corresponding $\bm{\beta}$.

To avoid restrictions binding ``early'' by mistake or by random change, the default setting in the accompanying software is to ``clip'' $E[1/\tau^2_k]$ at some large value (e.g. $10^{6}$) for the first few iterations. Further, the algorithm is initialized such that no restrictions are binding.

\subsection{Adaptive LASSO}

I sometimes rely on an adaptive LASSO in the spirit of \citet{gertheiss2010sparse} where the penalty is up-weighted for particular restrictions based on a consistent estimator of the difference. I also normalize each strength by the weights suggested in \citet{gertheiss2010sparse} to account for variables of different size. Both changes result in only a slight modification of the above. Assuming all of the weights are positive, this is equivalent to left multiplying $\bm{D}$ by an invertible diagonal matrix and thus all theoretical results apply automatically. 

\section{Details on Simulations}
\label{app:hetsims}

This section outlines more detailed results on the simulations in Section~\ref{section:het_sims}. First, to ensure comparability, all treatment effects shown are calculated using Monte Carlo integration. I estimate the effect of $d_i$ (moving from zero to one) for each unit $g$, marginalizing over $x_i$ using Monte Carlo integration. I draw 1,000 observations from a standard normal distribution: $\{\tilde{x}_i\}_{i=1}^{1000}$. For each group $g$, I calculate the estimated effect for each observation $i$: $E[y_i | d_i = 1, g[i] = g] - E[y_i | d_i = 0, g[i] = g]$. Averaging those together gets the estimate for each method and group. This allows for comparability across all proposed models and outcomes.

Next, I enumerates the methods used with reference to the specific R packages and formulae.

\begin{itemize}
	\item SSp - Structured Sparsity. This model is estimated using an agnostic (fully connected) structure with a LASSO penalty. I fit models over an equally spaced grid of $\lambda$ on the logarithmic scale. I choose the best model using the AIC with the degrees of freedom measure in \citet{tibshirani2012df}. Weights from \citet{gertheiss2010sparse} are used. I control for group-level effects in the simulations with a random effect estimated by approximate variational inference.
	\item A-SSp - Adaptive Structured Sparsity. The above model is fit with adaptive weights scaling each $K$ restriction using a ridge-stabilized estimate of the consistent model.
	\item FE - Fixed Effects. Estimated using \texttt{glm} and an interaction of indicator variables for each group with the treatment, i.e. \verb|glm(y ~ x + d * g)|.
	\item RE - Random Effects. Estimated using \texttt{glmer} \citep{bates2015lmer} and a random slope for the effect of treatment, i.e. \verb`glmer(y ~ x + d + (d | g))`
	\item BayesGLM - From \texttt{arm}; a generalized linear model with a prior on each coefficient from \cite{gelman2008weakly} to avoid separation. The formula is \verb|bayesglm(y ~ x + d * g)|
	\item LASSO - $\lambda$ chosen using 10-fold cross-validation from \texttt{glmnet} \citep{friedman2010cyclical}; the formula is \verb|cv.glmnet(y ~ x + d * g)|. It is weighted such that there is no penalization on the main treatment effect and thus a model with maximial sparsity recovers a generalized linear model with $x_i$ and $d_i$ included additively.
	\item ENet1 - Elastic Net with $\alpha = 0.50$. $\lambda$ chosen using 10-fold cross-validation from \texttt{glmnet}. Same formula as LASSO.
	\item ENet2 - Elastic Net with $\alpha = 0.25$. $\lambda$ chosen using 10-fold cross-validation from \texttt{glmnet}. Same formula as LASSO.
	\item FindIt - Estimated using default settings in \citet{imai2013findit}.
	\item SVM - Estimated using polynomial kernel and package \texttt{e1071} \citep{meyer2019svm}. Default settings were used. Note that a different package, \citet{hornik2009weka}, is used for the ensemble analysis.
	\item BART - Estimated using default settings from \texttt{BART} package (\cite{sparapani2019bart}).
	\item RF - Estimated using a forest of 1,000 trees with $\floor{G/3} + 2$ variables drawn per tree. Estimated using the \texttt{randomForest} package (\cite{liaw2002randomforest}). The standard design matrix (i.e. model matrix applied to \verb|y ~ X + treated + cgroup|) is provided.
	\item No Hetero. - No Heterogeneity. A model estimated with no heterogeneous effects, i.e. \verb|glm(y ~ x + d)|.
\end{itemize}

Next, I discuss tree-based methods and how they include categorical predictors. Rather than including the group identifiers as indicator variables---as in all other models, some random forest approaches (e.g. \cite{liaw2002randomforest}) adopt different strategies. For example, one approach is to order the categories at each split based on the observed outcome and use that \emph{ordered} variable to partition the groups (see, e.g., \cite[p. 310]{hastie2009elements}). 

There are two concerns with this approach; first, theoretically, it is unclear whether this is appropriate for an inferential (i.e. non-predictive) task as it, in some sense, uses the outcome to create variables to help predict the outcome! It is thus, in some sense, an ``unfair'' model to compare against. Second, existing software (\cite{liaw2002randomforest}) cannot include categorical variables with an arbitrary number of levels with certain non-standard ways of treating those variables.  I thus report results from the ``fair'' random forest where a design matrix of indicator variables is provided.

Preliminary experiments showed that alternative methods of including the categorical predictor resulted in better performance. I replicated simulations with a linear outcome and mostly grouped predictors where I ordered the factor based on the observed response before giving it to the estimation algorithm. This resulted in markedly better performance for the random forest, although it still was handedly beaten by LASSO, random effects and structured sparsity. Exploring this in detail across different choices of software and non-standard ways of treating categorical variables is reserved for future research.

\begin{figure}[!ht]
	\caption{Simulations for All Methods: Linear Data Generating Process}
	\label{fig:app_hetsims_linear}
	\includegraphics[width=\textwidth]{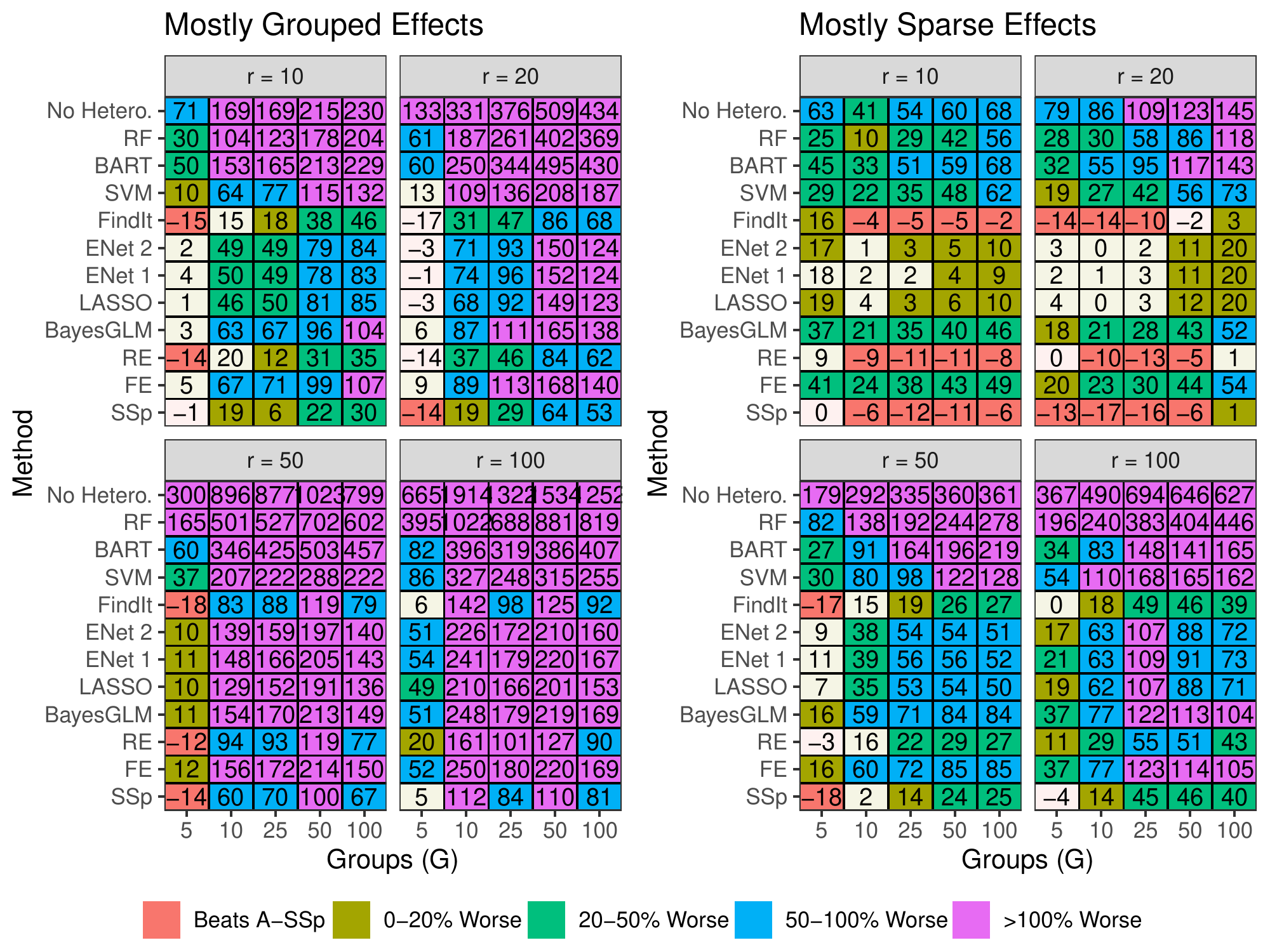}
	\caption*{\footnotesize \emph{Note}: The percentage change in RMSE vs A-SSp (adaptive structured sparsity) is shown: $\left(\mathrm{RMSE}_{k} - \mathrm{RMSE}_{\mathrm{A-SSp}}\right)/\mathrm{RMSE}_{\mathrm{A-SSp}} \cdot 100$. A positive number thus indicates \emph{worse} performance. All blocks that are lightly shaded indicate a difference that is \emph{not} statistically distinguishable at the 95\% level. The number reported is averaged across 100 simulations. For example, with $r = 10$, mostly grouped effects, $G = 5$, A-SSp beats an SVM by 10\% and loses to FindIt by 15\% (although the latter is not distinguishable from zero).}
\end{figure}

Third, Figure~\ref{fig:app_hetsims_linear} shows results by varying $G$ and $r$. To create interpretable results, I show the percent difference in RMSE (averaged across simulations) over adaptive Structured Sparsity (A-SSp): $\left(\mathrm{RMSE}_{k} - \mathrm{RMSE}_{\mathrm{A-SSp}}\right)/\mathrm{RMSE}_{\mathrm{A-SSp}} \cdot 100$. If the difference is statistically distinguishable at the 95\% level, the cell is \emph{solid}, otherwise it is light shaded.

Focusing on mostly grouped effects, when $G > 5$, structured sparsity always the best performing method and is almost always statistically distinguishably better.  Note the margin of improvement is rather large; the RMSE of alternative methods is often at least 50\% or over 100\% worse. Similarly, it out-performs competitors even when the truth is sparse in most settings when $G > 5$. It is beaten for small $r$ (e.g. $r = 10$, $r = 20$) by random effects, FindIt and the non-adaptive SSp in a distinguishable way, although for large $r$ it remains the dominant method. It is worth noting that the margin of improvement is noticeably smaller, although still considerable---often around 20-50\%.

\begin{figure}[!ht]
	\caption{Simulations for All Methods: Multinomial Data Generating Process}
	\label{fig:app_hetsims_multi}
	\includegraphics[width=\textwidth]{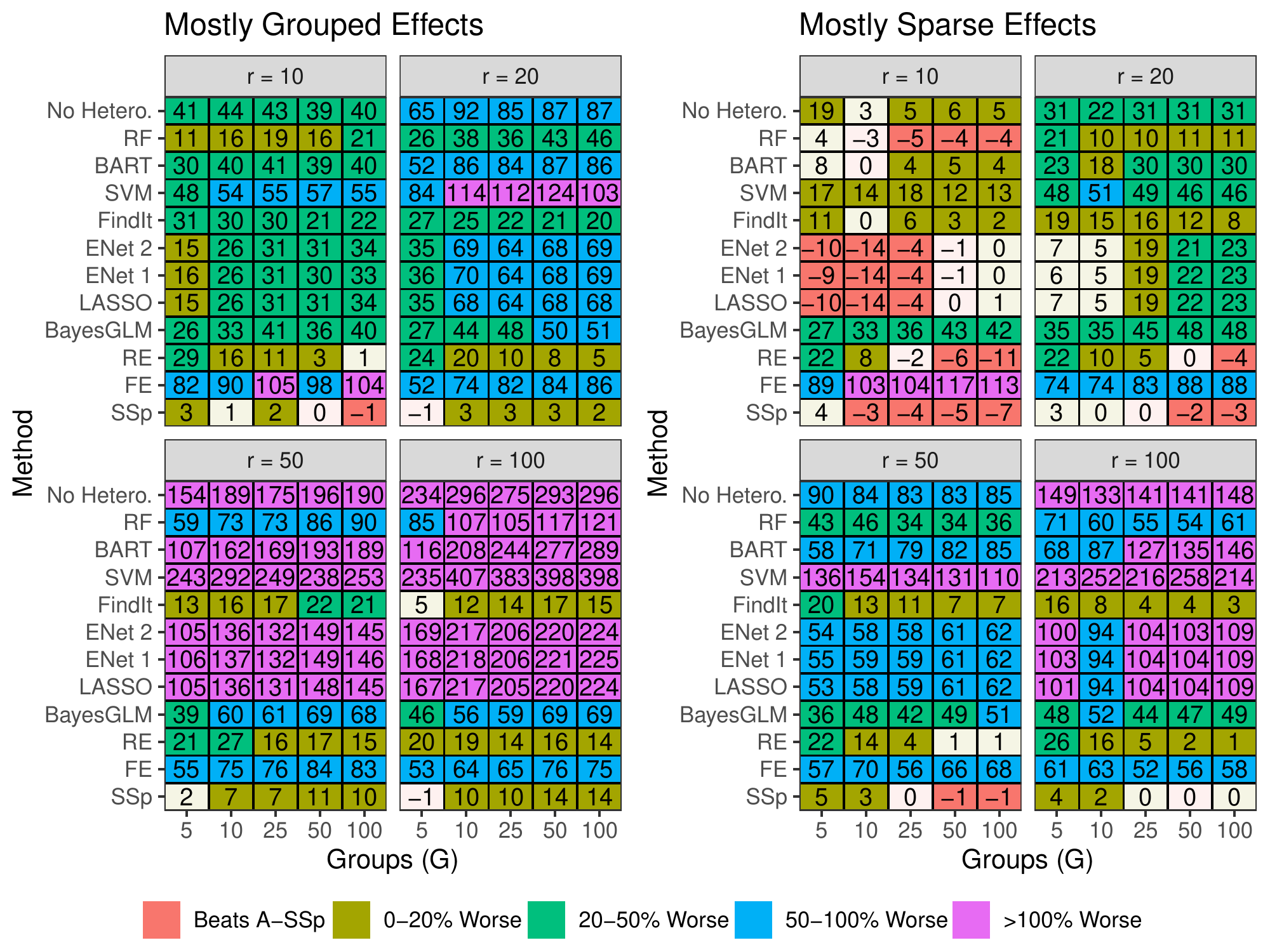}
	\caption*{\footnotesize \emph{Note}: The percentage change in RMSE vs A-SSp (adaptive structured sparsity) is shown: $\left(\mathrm{RMSE}_{k} - \mathrm{RMSE}_{\mathrm{A-SSp}}\right)/\mathrm{RMSE}_{\mathrm{A-SSp}} \cdot 100$. A positive number thus indicates \emph{worse} performance. All blocks that are lightly shaded indicate a difference that is \emph{not} statistically distinguishable at the 95\% level. The number reported is averaged across 100 simulations. For example, with $r = 10$, mostly grouped effects, $G = 5$, A-SSp beats an SVM by 48\% and beats FindIt by 31\%; both are distinguishable from zero.}
\end{figure}

Finally, for the models with a binary outcome, the results are shown in Figure~\ref{fig:app_hetsims_multi}. The generative model is as follows:

$$y_i \sim \mathrm{Bernoulli}\left(\frac{\exp(x_i + \tau_{g[i]} d_i)}{1+\exp(x_i + \tau_{g[i]} d_i)}\right)$$

For the mostly grouped case ($\mathcal{S} = \floor{G/2}$), structured sparsity again out-performs almost all methods when $G > 5$. Note, however, that the magnitude of improvement is usually smaller than in the linear case although almost always statistically distinguishable; the magnitude of the improvement grows as $r$ grows. In the mostly sparse case, the results are more mixed for structured sparsity especially when $r = 10$, but the improvements are still usually distinguishable from competitor methods although the magnitude is smaller than in the mostly grouped case---as in the linear model.

Consider an alternative summarization: Of the 80 simulation environments examined, adaptive structured sparsity performs the best in terms of mean RMSE across simulations in 43 environments and non-adaptive performs the best in 20 environments. The next best method (FindIt) performs best in 7, usually when $G = 5$. Adaptive and non-adaptive structured sparsity are in the top three methods in 60 and 74, respectively, of the environments with the next closest method (random effects) at 47. 

A final remark is that non-adaptive structured sparsity appears to out-perform structured sparsity when the truth is mostly sparse (wining 25/40 times vs. 9/40).

\section{Details on Credit-Claiming Analysis}
\label{app:grimmer}

\subsection{Bayesian Convergence Diagnostics}

\begin{figure}[!ht]
	\caption{Convergence Diagnostics on $\bm{\beta}$}
	\begin{subfigure}[b]{0.5\textwidth}
		\caption{Gelman-Rubin Convergence Statistics}
		\label{fig:gelman_rubin}
		\includegraphics[width=\textwidth]{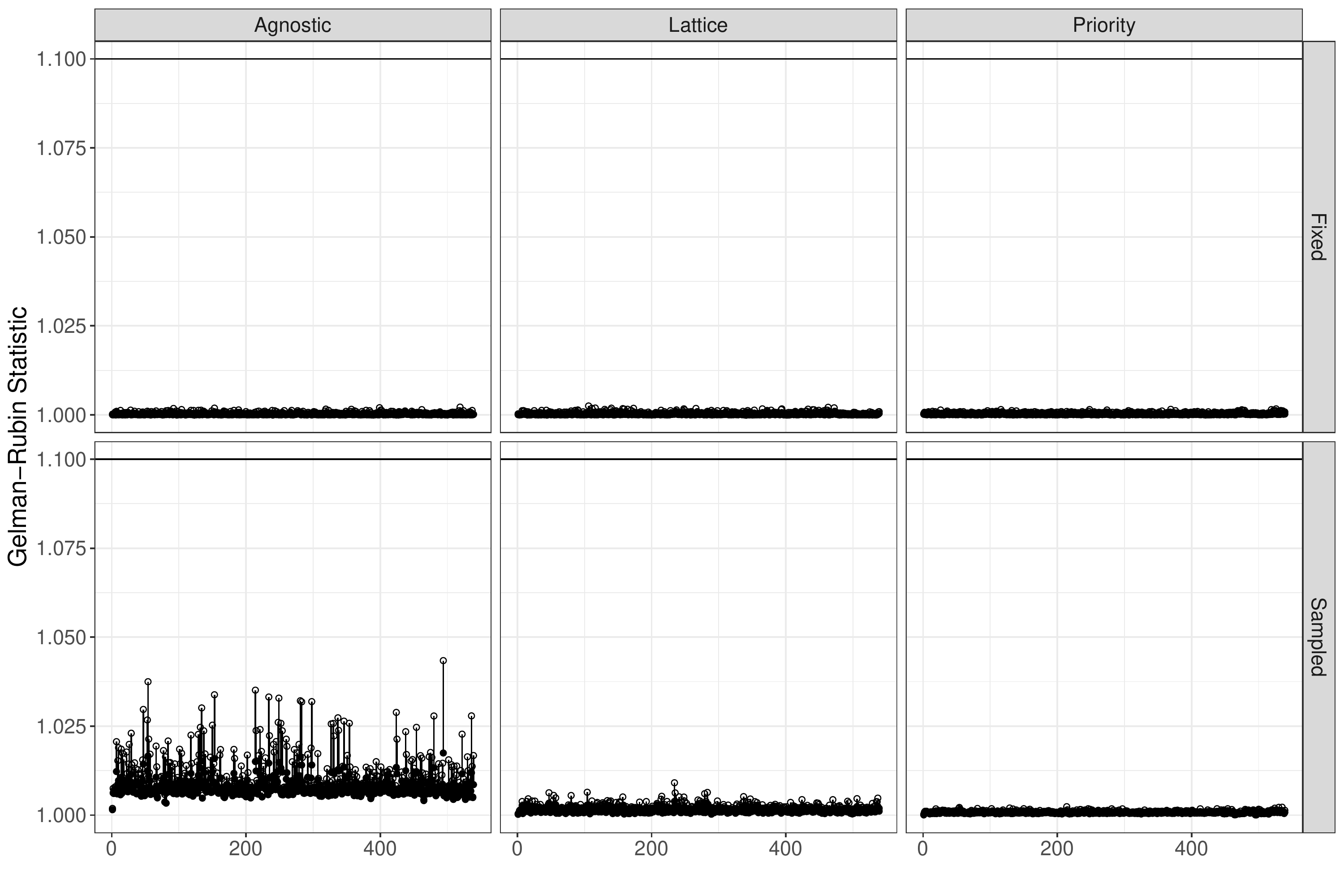}
	\end{subfigure}
	\begin{subfigure}[b]{0.5\textwidth}
		\caption{Geweke Convergence Statistics}	\includegraphics[width=\textwidth]{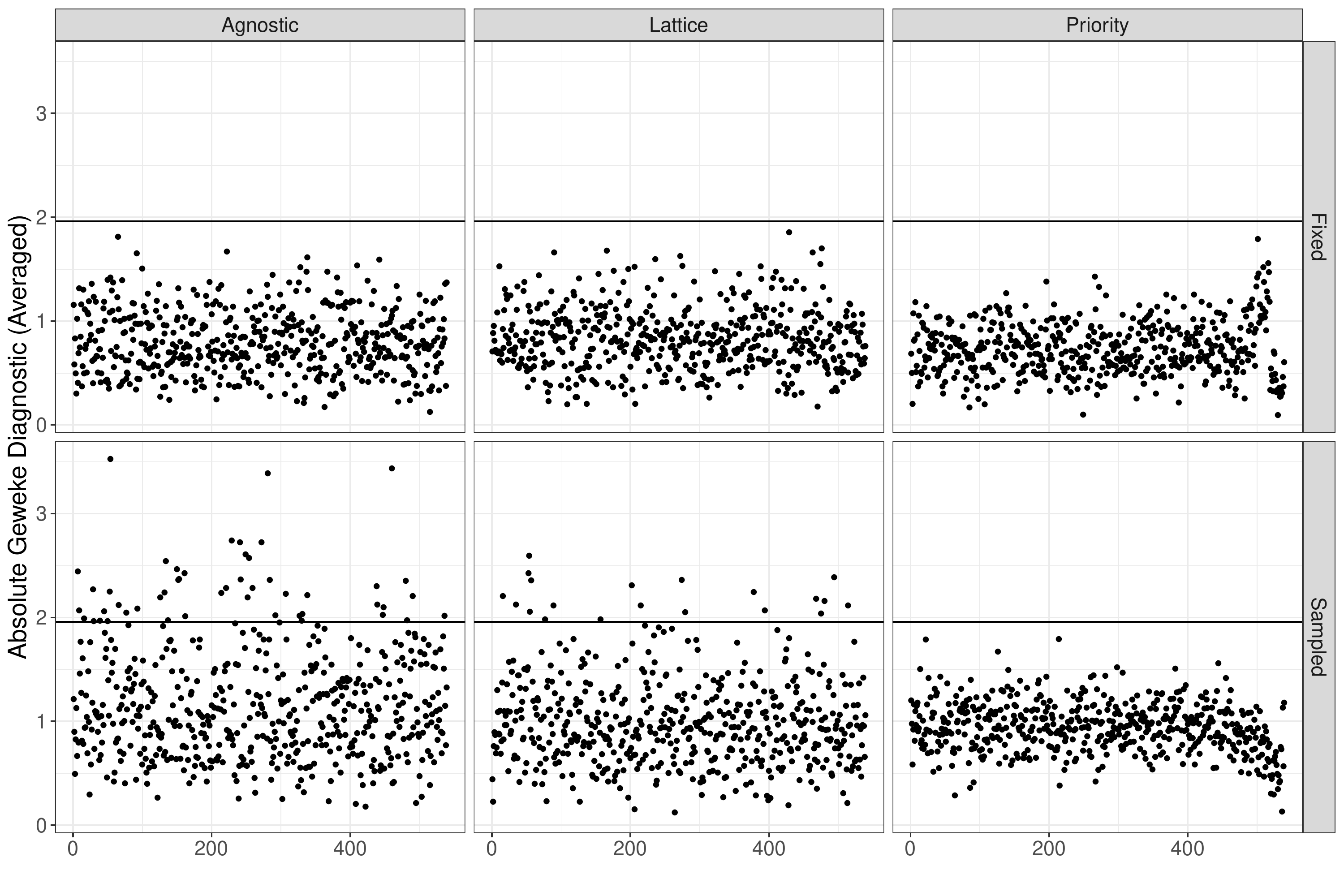}
	\end{subfigure}
\end{figure}

I present results on the convergence of the posterior sampler for the three structured sparse models outlined in the main text. I ran each model with over-dispersed starting values (drawing from a uniform ranging from -3 to 3) for 4 chains, 10,000 iterations each and discarded the first 5,000 as burn-in. This gives 20,000 samples from the posterior. Figure~\ref{fig:gelman_rubin} reports the Gelman-Rubin statistic for all parameters in the model. A threshold of 1.1 is a common test for convergence. All parameters are below this value as are the upper confidence statistic reported by \texttt{coda}.

I also report the Geweke statistic. To summarize across chains, I report the average (absolute) statistic across all four chains. Looking at each statistic individually, it is above the 1.96 threshold in below 5\% of parameter-chain combinations when $\lambda$ is fixed. When $\lambda$ is sampled, it is higher; around 12\% for the lattice model used in the main text. 

\begin{figure}[!ht]
	\caption{Lambda Trace Plot}
	\label{fig:trace_lambda}
	\includegraphics[width=\textwidth]{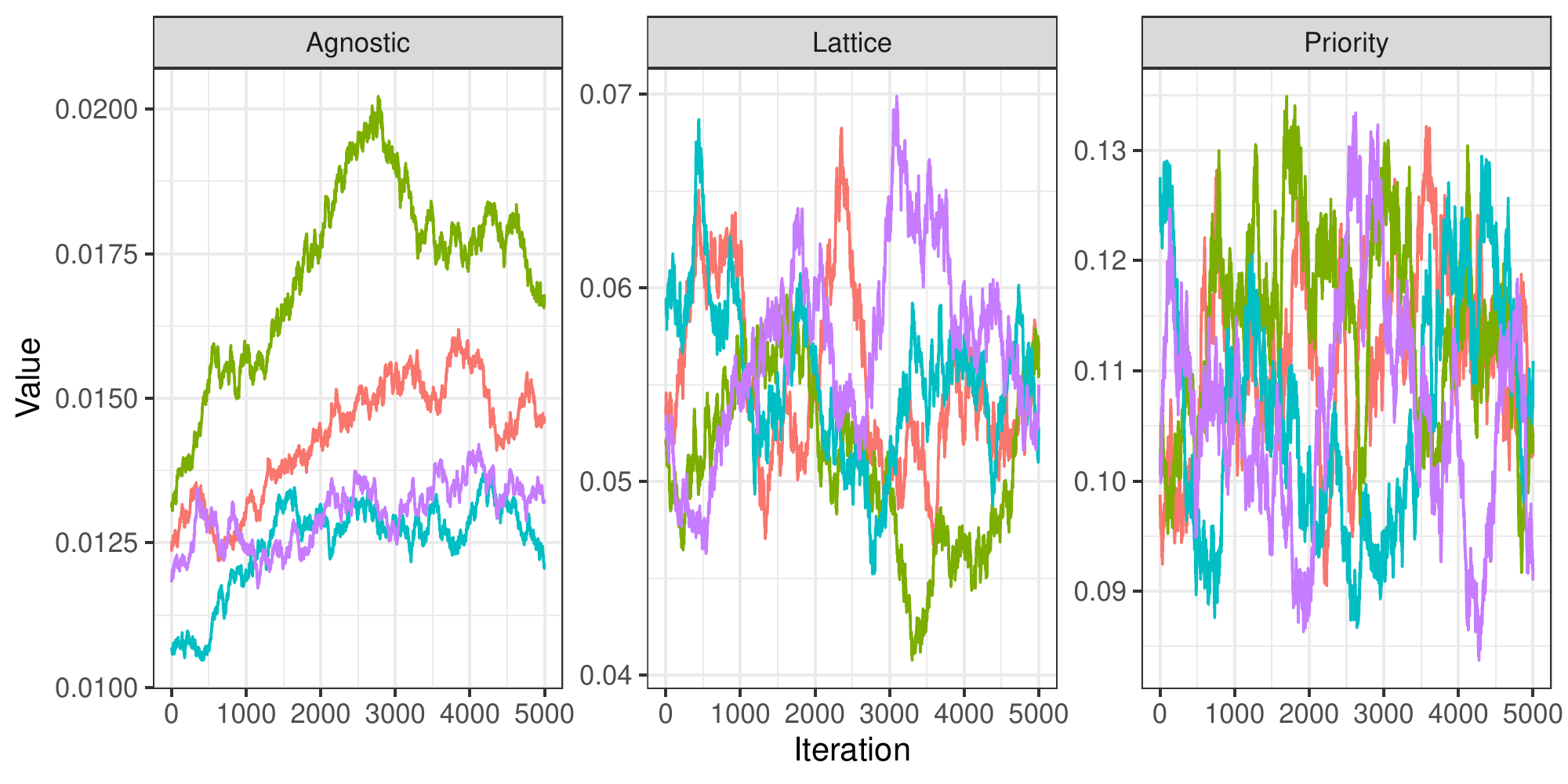}
\end{figure}

The likely cause of this poorer behavior is the issues with the convergence of $\lambda$ when sampled for the lattice and agnostic structure. Even after considerable burn-in, mixing appears to be slow as Figure~\ref{fig:trace_lambda} shows---especially for the agnostic structure. A different sampler may perform better. Exploring this is an important area for future research. Fortunately, as the results in the next sub-section show, fixing $\lambda$ at $\lambda^*$ returns nearly identical estimates suggesting that the lack of stationarity in $\lambda$ will not materially undermine the results.

I attempted to fix this problem by running a much longer model for the lattice structure: 20,000 iterations after 20,000 of burn-in on four chains. This improves within-chain mixing (i.e. the Geweke diagnostic for all chains is below 1.96), but the Gelman-Rubin statistic is still stubbornly high at 1.17. 

\subsection{Substantive Additional Results}

\begin{figure}[!ht]
	\caption{Comparison of Structures}
	\label{fig:alt_struc}
	\includegraphics[width=\textwidth]{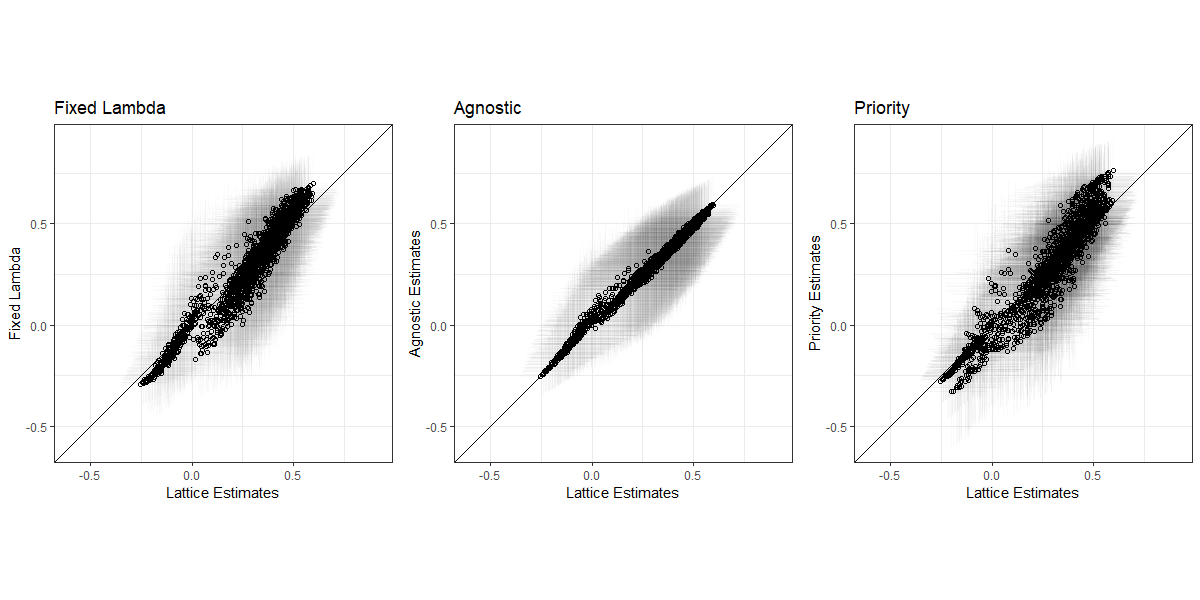}
	\caption*{\footnotesize \emph{Note}: Each figure plots the posterior median of one of three alterantive methods (lattice structure with fixed $\lambda$, agnostic structure, and priority structure) against the estimates with a lattice structure. The 90\% credible intervals for each parameter are shown in light grey below.}
\end{figure}

Figure~\ref{fig:alt_struc} shows the stability of treatment effects across structure. It plots the estimated treatment effects for all treatment/respondent combinations. The lattice effects (reported in the main text) are shown on the horizontal axis. Each panel reports a particular comparison; the model estimated with a fixed $\lambda^*$ at the optimal value found via the AIC, an agnostic structure and a priority structure.

Note the tight correlation between the estimates with a fixed $\lambda$ and lattice structure and allowing it to be sampled. This provides confidence that the results are not being driven by the lack of stationarity in the sampling of $\lambda$. The comparison with the agnostic method also suggests that the choice of structure may be less important if the limiting conditions are the same. The priority results look somewhat different but are closely correlated.

\end{document}